\documentclass{aa}

\usepackage{graphicx}
\usepackage{txfonts}
\usepackage{hyperref}			
\usepackage{subcaption}
\usepackage{multirow}
\usepackage{marvosym}
\usepackage[euler]{textgreek}

\begin{document}

\title{Expected performances of the Characterising Exoplanet Satellite (CHEOPS)}
\subtitle{I. Photometric performances from ground-based calibration}
\titlerunning{Expected performances of the Characterising Exoplanet Satellite (CHEOPS). I.}

\author{
	A. Deline\inst{1}\fnmsep\thanks{\email{adrien.deline@unige.ch}} \and
	D. Queloz\inst{1}\fnmsep\inst{2} \and
	B. Chazelas\inst{1} \and
	M. Sordet\inst{1} \and
	F. Wildi\inst{1} \and
	A. Fortier\inst{3} \and
	C. Broeg\inst{3} \and
	D. Futyan\inst{1} \and
	W. Benz\inst{3}
	}
\authorrunning{A. Deline et al.}

\institute{Department of Astronomy, University of Geneva, Chemin des Maillettes 51, 1290 Versoix, Switzerland
	\and
	Cavendish Laboratory, University of Cambridge, J J Thomson Avenue, Cambridge, CB3 0HE, UK
	\and
	Centre for Space and Habitability, University of Bern, Gesellschaftsstrasse 6, 3012 Bern, Switzerland
	}

\date{Received May 29, 2019; accepted August 5, 2019}

\abstract
	{Characterisation of Earth-size exoplanets through transit photometry stimulated new generations of high-precision instruments. In that respect, the \emph{Characterising Exoplanet Satellite} (CHEOPS) is designed to perform photometric observations of bright stars to obtain precise radii measurements of transiting planets. CHEOPS will have the capability to follow up bright hosts provided by radial-velocity facilities. With the recent launch of the \emph{Transiting Exoplanet Survey Satellite} (TESS), CHEOPS may also be able to confirm some of the long-period TESS candidates and to improve the radii precision of confirmed exoplanets. }
	{The high-precision photometry of CHEOPS relies on careful on-ground calibration of its payload. For that purpose, intensive pre-launch campaigns of measurements were carried out to calibrate the instrument and characterise its photometric performances. This work reports on main results of these campaigns. It provides a complete analysis of data sets and it estimates in-flight photometric performance by mean of end-to-end simulation. Instrumental systematics have been measured by carrying out long-term calibration sequences. Using end-to end model, we simulated transit observations to evaluate the impact of in-orbit behaviour of the satellite and to determine the achievable precision on the planetary radii measurement.}
	{After introducing key-results from the payload calibration, we focus on the data analysis of series of long-term measurements of uniformly illuminated images. The recorded frames are corrected for instrumental effects and a mean photometric signal is computed on each image. The resulting light curve is corrected for systematics related to laboratory temperature fluctuations. Transit observations are simulated, considering the payload performance parameters. The data are corrected using calibration results and estimates of the background level and position of the stellar image. The light curve is extracted using aperture photometry and analysed with a transit model using a Markov chain Monte Carlo algorithm.}
	{In our analysis, we show that the calibration test set-up induces thermally-correlated features in the data that can be corrected in post-processing to improve the quality of the light curves. We find that on-ground photometric performances of the instrument measured after this correction is of the order of 15~parts~per~million over 5~hours. Using our end-to-end simulation, one determines that measurements of planet-to-star radii ratio with a precision of 2\% for a Neptune-size planet transiting a K-dwarf star and 5\% for an Earth-size planet orbiting a Sun-like star are possible with CHEOPS. It corresponds to signal-to-noise ratios on the transit depths of 25 and 10 respectively, allowing the characterisation and detection of these planets. The pre-launch CHEOPS performances are shown to be compliant with the mission requirements.}
	{}

\keywords{planets and satellites: detection -- techniques: photometric -- space vehicles: instruments -- instrumentation: photometers}	

\maketitle

\section{Introduction} \label{sec:intro}
	
	After the first observations of a transiting exoplanet in 1999 \citep{charbonneau_hd209458b,henry_hd209458b}, the number of planets identified to pass in front of their host stars kept increasing. Following on early successes of wide field surveys like WASP \citep{pollacco_wasp}, the launch in 2009 of \emph{Kepler}, a dedicated transit search satellite \citep{koch_kepler,borucki_kepler}, produced a continuous stream of transiting planet candidates. At the end of its nominal duration, the \emph{Kepler} mission identified more than 4700 candidates, including over 2300 confirmed extra-solar planets in transit. The radii of these objects, derived from the in-transit light curve \citep{winn_transit}, span a large range with Mars-size up to Jupiter-size planets. Unfortunately, the vast majority of stars with transit planets identified by \emph{Kepler} are too faint to be easily followed up from the ground, making the determination of the planetary masses by precise Doppler measurements challenging. Combining masses and radii measurements of exoplanets is essential to obtain an estimate of their bulk densities and to retrieve information about their physical structures and possibly hints on their formation processes. The \emph{Transiting Exoplanet Survey Satellite} (TESS; \citealt{ricker_tess}) was launched in April 2018 with the purpose of performing a whole-sky transit search survey on brighter stars. Already a dozen of confirmed transiting exoplanets have been found \citep{gandolfi_tess,wang_tess,trifonov_tess}. Typical duration of observation sequence for each field observed by TESS is 27 days. Transiting planet with period longer than ten days will best transit once or twice, limiting the precision of planetary radius measurement for these systems. In this context, the \emph{Characterising Exoplanet Satellite} (CHEOPS; \mbox{\citealt{broeg_cheops,fortier_cheops}}) has been designed to perform pointed follow-up observations of bright stars with the goal to reach a photometric precision similar to \emph{Kepler}. When combined with TESS survey potential, CHEOPS will have the capability to confirm candidates and to improve some of these radii measurements.
	
	CHEOPS will be launch on a circular Sun-synchronous orbit at an altitude of 700\,km and a local time of ascending node of 6AM. This configuration corresponds to a nearly polar orbit (inclination of~98\degr) with a period of almost 100~minutes. CHEOPS will be able to point target up to 60\degr\ off the ecliptic plane corresponding to an overall sky coverage of 70\%. The cadence of the observations will be better than 1 minute. The nominal mission duration is 3.5~years, with a goal to 5~years. The launch date is planned between mid-October and mid-November~2019.
	
	The spacecraft is made of two main modules: the platform and the payload. The platform maintains the thermal stability of the payload, it ensures the powering of the instrument and it operates the data down-link to Earth. The performance of the payload attitude and control system (AOCS) is designed to maintain tracking on target with better than 4~arcseconds~(rms). The payload includes the telescope, a Ritchey-Chr\'etien design with a primary mirror of 32\,cm in diameter, and an efficient stray-light suppression system made of a baffle and a field stop. The payload detector is a 1024$\times$1024-pixel frame-transfer back-illuminated charge-coupled device (CCD) from the CCD47-20 family of sensors manufactured by the company \emph{e2v}. The CCD and the front-end electronics are both thermally stabilised at the precision of 10\,mK, with operating temperatures of -40\degr C and -10\degr C respectively, in order to limit the noise contributions of the dark current and electronic gain variability. The pixel scale on the detector corresponds to 1~arcsecond on the sky.
	
\begin{figure}
		\centering
		\includegraphics[width=.9\hsize,trim={.5cm 0cm 1cm 1.5cm},clip]{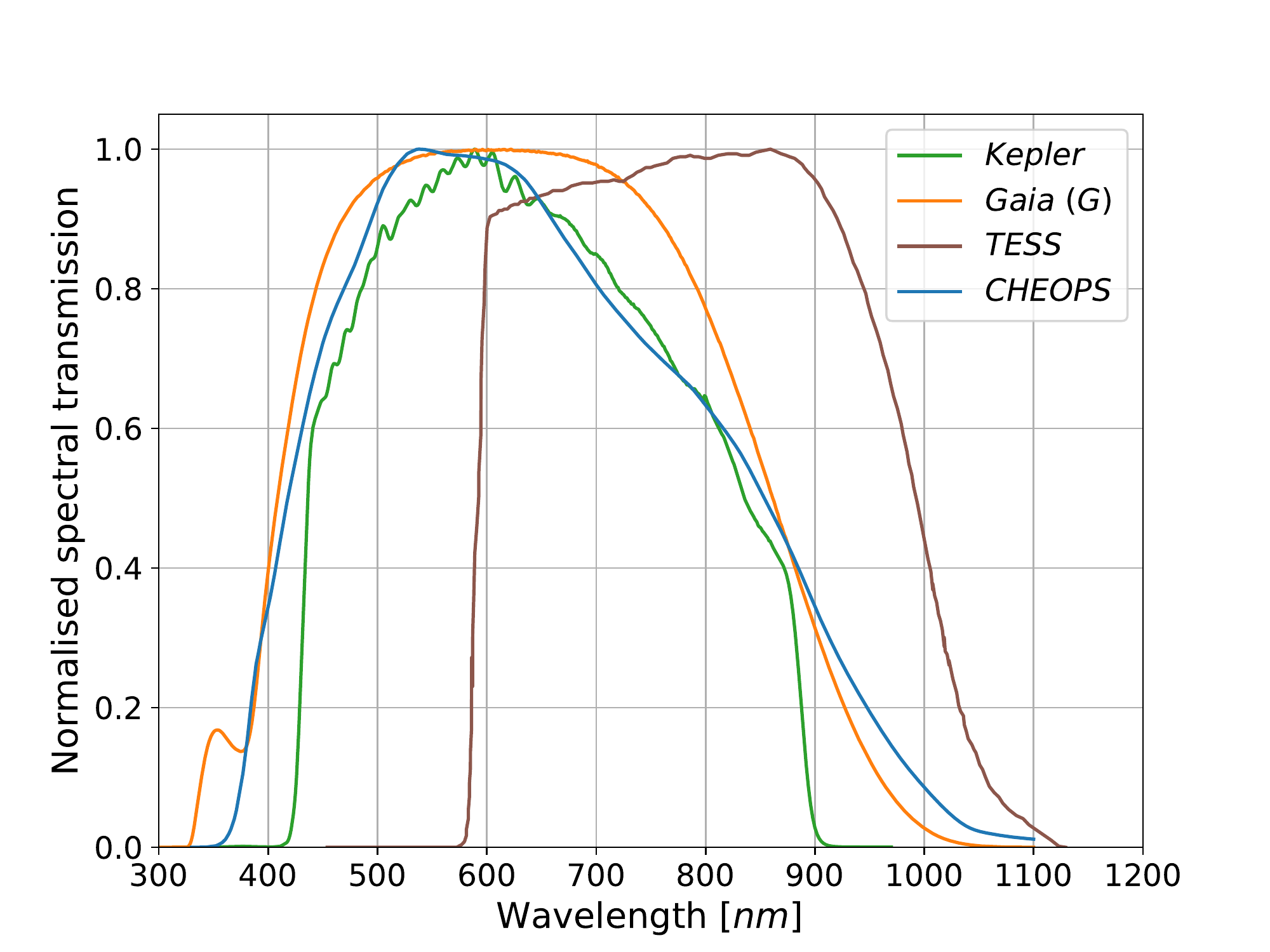}
		\caption{Normalised spectral transmissions (in units of energy) of CHEOPS (\emph{blue}) and other space missions: \emph{Kepler} in \emph{green} (\emph{Kepler Instrument Handbook}; \citealt{van_cleve_kepler}), \emph{Gaia} in \emph{orange} (G passband; \citealt{maiz_gaia}) and TESS in \emph{brown} \citep{ricker_tess}.}
		\label{fig:passbands}
	\end{figure}

	The point spread function (PSF) of target image on the detector has a radius of 12~pixels (distance from the PSF centre at which 90\% of the energy is encircled), to ensure illuminating enough pixels to minimise the photometric effect of satellite tracking residuals. To maximise science data down-link and to achieve 1-minute sample rate, a circular sub-frame of 200$\times$200~pixels centred on the target is downloaded. It represents a field of view of 3.3\,arcmin in diameter. The photometric spectral domain of CHEOPS covers a range from 330\,nm to 1100\,nm (see Fig.~\ref{fig:passbands}), similarly to the space mission \emph{Gaia} \citep{gaia}. During the course of its orbit around the Earth, the spacecraft keeps rolling to maintain its cold plate radiators opposite to the Earth direction. As a result, the field of view rotates around the pointing line of sight defined by the target location on the detector. This creates a circular motion of the background stars at a rate of one rotation per orbital period (100~minutes).
	
	CHEOPS has been designed with two science requirements in mind. First the capability to detect an Earth-size planet with a period of 50~days transiting a G5~star ($R_*\,$=$\,0.9\,R_\odot$) with \mbox{V-magnitude} of 9. This corresponds to a light curve with a typical noise of 20~parts per million (ppm) when 6~hours of data are combined. Second, to be able to measure a transit light curve with signal-to-noise ratio (SNR) of~30 on a K-type star ($R_*\,$=$\,0.7\,R_\odot$) from a transiting Neptune-size planet with a period of 13~days. This corresponds to reaching a photometric precision of 85\,ppm on the light curve in 3~hours.
	
	This work is based on results obtained during ground calibration campaigns to assess the photometric precision of CHEOPS satellite. A brief description of equipment used is mentioned in section~\ref{sec:bench}. Readers interested to learn more about it will find relevant references in the section. The main calibration products required for the data analysis of this study are described in (section~\ref{sec:ccd}). We will then develop two approaches to assess CHEOPS performances. The first based on the analysis of long-term data set obtained during the calibration campaign (section~\ref{sec:cal_phot}). The second by developing an end-to-end data simulation with transit observations (section~\ref{sec:sim}).
	
	This paper is part of a mini-series entitled \emph{Expected performances of the Characterising Exoplanet Satellite (CHEOPS)} that includes two other publications. One describes the software \texttt{CHEOPSim} developed for the CHEOPS mission to simulate scientific data (\emph{The CHEOPS simulator}; \citealt{futyan_cheopsim}). \texttt{CHEOPSim} is briefly introduced in section~\ref{sssec:cheopsim}. The other publication details the data reduction pipeline~(DRP) that will be used during the mission to extract the photometry from the raw data and provide light curves of planetary systems (\emph{The CHEOPS data reduction pipeline: architecture and performances}; \citealt{hoyer_cheops-drp}).

\section{The calibration test set-up} \label{sec:bench}
	
\begin{figure}
		\centering
		\includegraphics[width=\hsize]{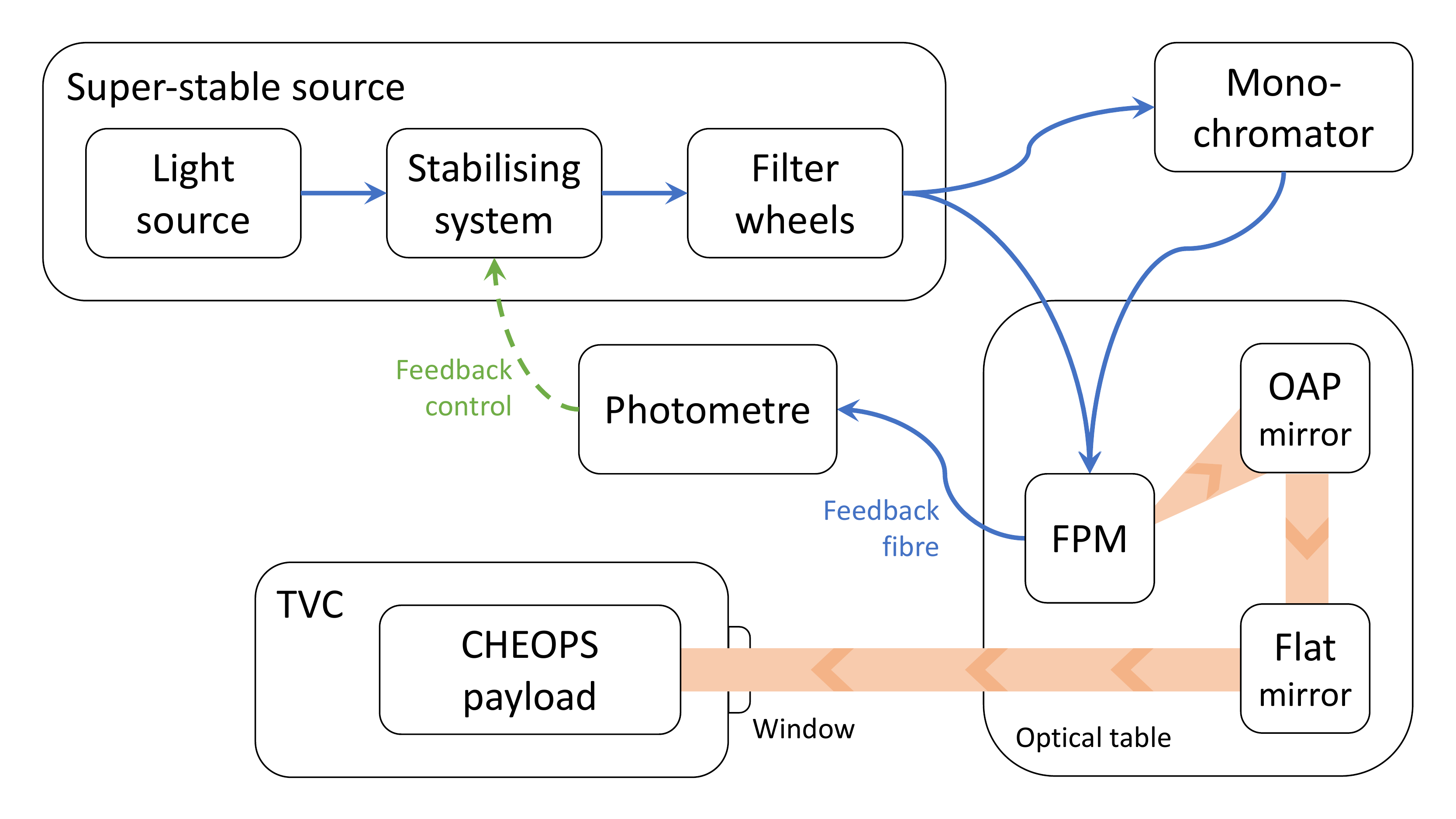}
		\caption{Functional diagram of the CHEOPS calibration bench. The \emph{blue} arrows represent the optical fibres used to guide the light through the calibration bench. The super-stable source is equipped with a system that can send the light in the monochromator or bypass it. The light beam (\emph{orange}) emitted by the focal plane module (FPM) first hits the off-axis parabolic (OAP) mirror before being folded by a flat tip-tilt mirror. After exiting the bench, the light goes through a window into the thermal-vacuum chamber (TVC) and enters the payload telescope. In the FPM, part of the source flux is picked up and sent to a feedback photometer that is used to control the stabilising system of the Super-Stable Source.}
		\label{fig:cal_bench}
	\end{figure}

	A set-up was specifically developed and integrated to perform the calibration of CHEOPS payload, including the flying CCD \citep{wildi_bench}. The system is based on a "super-stable (light) source" (SSS) capable of reaching a stability of 3\,ppm over 1~minute \citep{wildi_sss} and covering the whole passband of CHEOPS. The SSS system could also be used to modify the source spectrum by inserting optical filters or by directing the light through a monochromator. The stabilised flux was then injected in an optical fibre and guided to the Focal Plane Module (FPM) of the set-up, located on an optical table. The FPM could switch between two different modes: a point-source mode simulating a single-star field and an extended-source mode illuminating uniformly (better than 99.5\% uniformity) the whole CHEOPS CCD. The diverging beam coming from the FPM was collimated by an off-axis parabolic mirror before being directed towards the payload by a tip-tilt folding flat mirror. For performance tests, the payload was located inside a Thermal-Vacuum Chamber (TVC) with a window allowing the pupil of the telescope to be fully illuminated by the stabilised beam (see Fig.~\ref{fig:cal_bench}). The purpose of the TVC was to reproduce the space conditions in which the satellite will have to operate in orbit. In that respect, the test set-up provides the possibility to vary some of the payload parameters allowing us to explore the ranges within which we expect CHEOPS to evolve while in space.
	
	During the calibration campaign, technical limitations and lack of time prevented us to perform long and systematic series of precise photometric measurements with star-like point-source images. This mode was only used to measure the shape of the PSF on the detector, essential to produce a realistic end-to-end modelling. We conducted our series of photometric measurements using the extended-source mode and all results presented in this work were obtained with uniformly illuminated images. During the course of this work, we found out that the light source stabilisation was correlated with the laboratory temperature. The optical fibre used by the feedback loop has a refractive index sensitive to thermal changes \citep{moraleda_fiber,wang_fiber}. This means that the flux measured by the photometer to drive the lamp stabilisation mechanism is inaccurate. This leads to erroneous feedback stabilisation commands. Unfortunately, time pressure to deliver the payload for integration on the platform prevented us to elaborate a solution to fix the feedback loop. We decided to operate the whole system in open loop, keeping the photometer only as a flux sensor, and to perform post-processing correction of this effect on the recorded data.

\section{Instrument calibration products} \label{sec:ccd}
	
	In this section, we essentially focus on calibration measurements relevant for high-precision photometry, assuming nominal operations. We briefly present calibration results and stability performances of electronic bias offset, detector dark current and read-out electronic system gain. The process to measure the photo-response non-uniformity of the detector is presented. The precision of flat-field correction is finally assessed.

	\subsection{Bias offset, dark current, gain and instrument stability} \label{ssec:ccd_perf}
		
		The bias level was computed with an average precision of 0.2\,ADU/pixel. The read-out noise (RON) is 14\,e$^-$ and 7\,e$^-$ for respectively read-out frequencies of 230\,kHz and 100\,kHz. When operated at nominal temperature, the CCD dark current measured over the whole detector frame has an average value of~0.028\,e$^-$/s (value of the \emph{mode} of the intensity histogram). We identified two hot pixels ($>10\,\text{e}^-$/s) and five warm pixels ($>5\,\text{e}^-$/s) on the detector.
		
		The read-out electronic system gain and the non-linearity are two quantities describing the conversion of the number of electrons recorded by the CCD to digital units stored by the computer. The scaling conversion is the gain, a multiplicative factor expressed in ADU/e$^-$. It depends mainly on the temperature and power voltages operating the detector. The non-linearity term quantifies the deviation from the gain conversion above and mostly depends on the CCD read-out frequency. The value of the gain in nominal conditions was measured with a precision of~0.6\%. The gain variations during changes of CCD temperature and power operating voltages were measured and modelled during the characterisation phases of the detector \citep{deline_ccd}. Measured gain sensitivities are typically of the order of~30\,ppm/mV and 1\,ppm/mK for the voltages and temperature respectively. Using a simple thermal model (a second-order spline curve) deviations are found smaller than 0.1\%\,rms. During our measurements, we stayed within dynamical range of our detector defined when the non-linearity is less than 3\%. This corresponds to an upper value of 121\,ke$^-$.
		
		Long series of measurements, over hours, have shown bias level and read-out noise to be very stable with variations smaller than 30\,mADU/day and 1\,mADU/day respectively. During these sequences the stability of the bias voltages powering the detector was of the order of tens of \textmu V, whereas the temperatures were varying on the mK scale. When combining these numbers with the gain sensitivities, we find that for nominal operations the effect of the gain variations is of the order of 1\,ppm and have a negligible contribution to our noise budget.

	\subsection{Flat fields} \label{ssec:ff}
		
		Due to the manufacturing process, CCD have small spatial variations in thickness, pixel sizes, well depths and substrate material. These imperfections generate pixel-to-pixel non-uniformity of detector photo-response. If not accurately measured, when combined with the expected guiding errors of a few arcseconds, it creates photometric errors. Accurate characterisation of pixel photo-response non-uniformity, or flat field, is an essential step to obtain precise photometry.

		\subsubsection{Measurements} \label{sssec:ff_meas}
			
\begin{figure}
				\centering
				\includegraphics[width=\hsize,trim={3.2cm 1.9cm 4.5cm 2.2cm},clip]{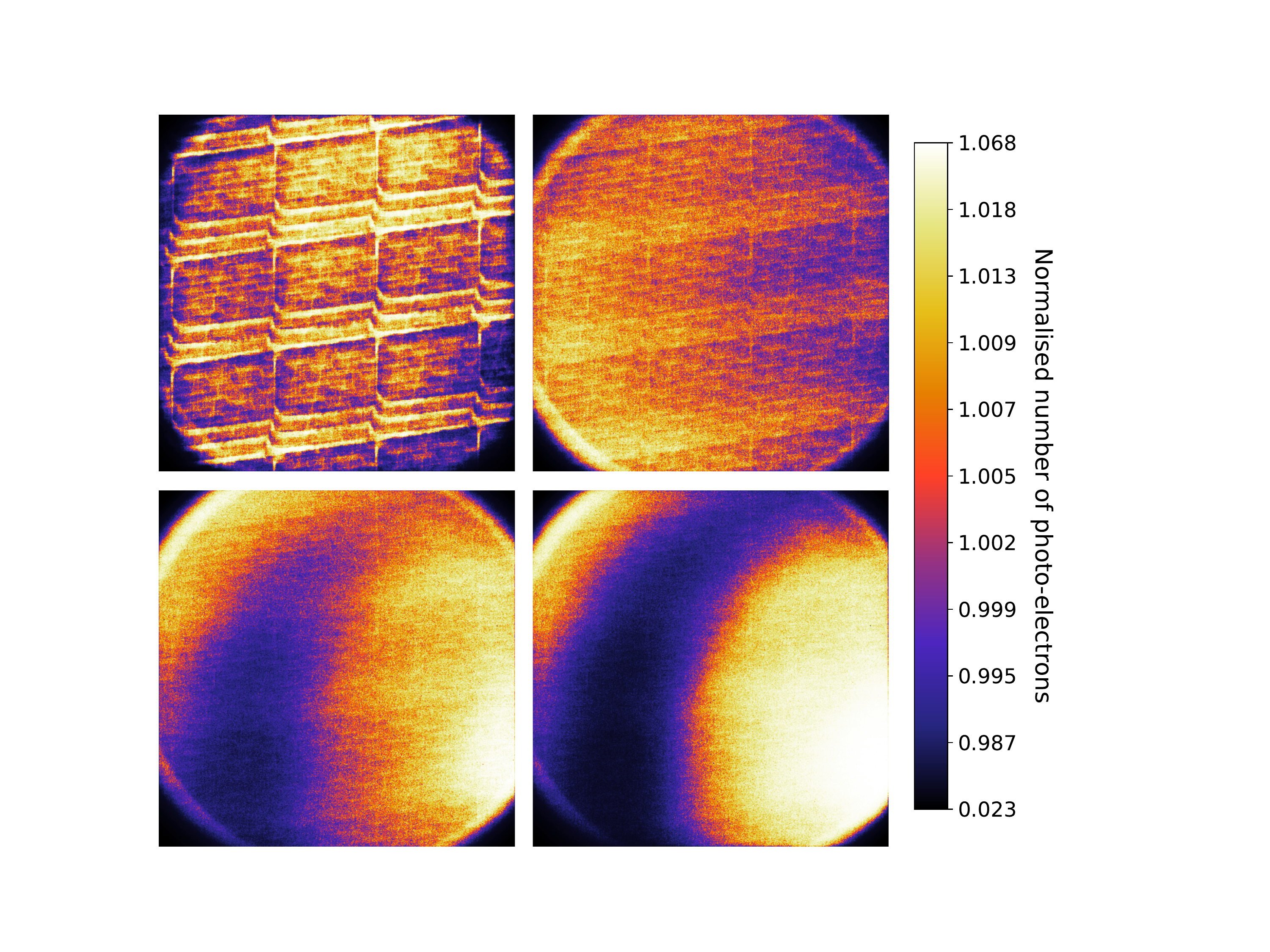}
				\caption{Flat fields measured with the narrow-band filters at 442.5\,nm (\emph{upper left}), 580.3\,nm (\emph{upper right}) and 734.3\,nm (\emph{lower left}) and the broadband Johnson/Bessel I filter (\emph{lower right}). The colour scale is not linear and has been optimised to highlight small variations using histogram equalisation. The four dark corners on all images are not dead pixels but are caused by a stray-light suppression component of the payload called \emph{field~stop}.}
				\label{fig:ff}
			\end{figure}

			Photons are absorbed by CCD at different depths depending on their wavelengths. This colour dependency has to be accounted to accurately measured detector flat-field responses since the spectral distribution of the stellar flux depends on the nature of the star considered. To characterise in details the detector response over the CHEOPS passband, we measured flat fields on restricted wavelength domain using 23 different narrow-band filters ($\Delta\text{\textlambda}=30$\,nm) from 442.5\,nm to 734.3\,nm and 4 broadband Johnson/Bessel filters (U, B, R, I) (see Fig.~\ref{fig:ff}). Flat-field images are calibrated, converted in photo-electrons and normalised to their mean. The average rms~precision per pixel for each flat field is $6\,10^{-4}$. Different spatial features are visible for each range of wavelength, such as the typical \emph{diamond pattern} visible in the blue. The measurements were repeated a week later and we could not detect any variation in the flat-field images, suggesting flat fields are stable on timescales of days.
			
		\subsubsection{Flat-field synthesis} \label{sssec:ff_syn}
			
\begin{figure}
				\centering
				\includegraphics[width=\hsize,trim={1.4cm 1cm .7cm 2.3cm},clip]{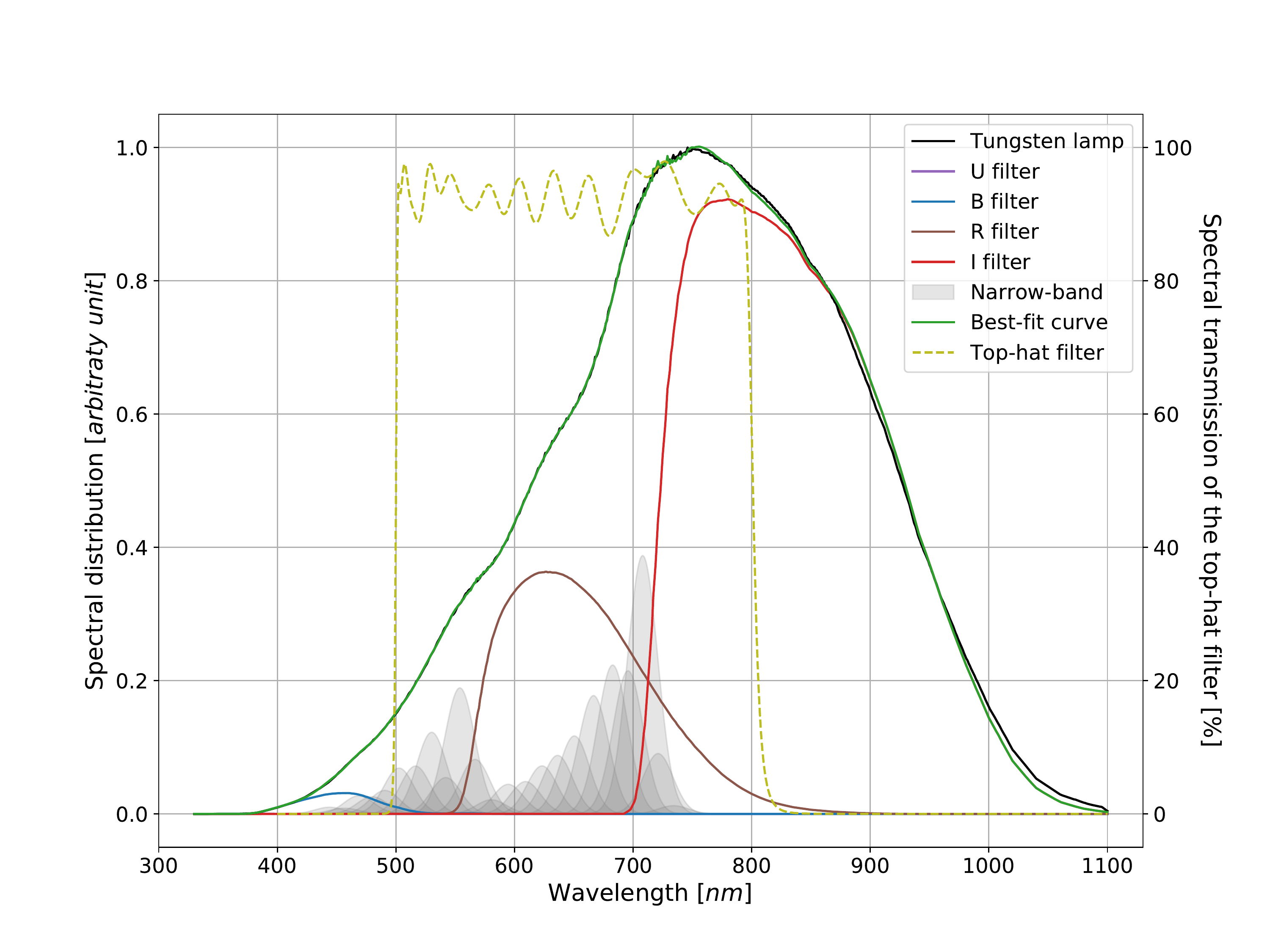}
				\caption{Weighted spectral distributions of the U,~B,~R,~I (Johnson/Bessel filters) and narrow-band filters used to fit the Tungsten lamp profile (\emph{black}). The weighted sum of the distributions is the "Best-fit curve" (\emph{green}) . The spectral transmission of the top-hat filter (\emph{yellow dashed line}) has been added for information (see section~\ref{ssec:cal_meas}).}
				\label{fig:ffc_sp}
			\end{figure}

\begin{figure}
				\centering
				\includegraphics[width=.65\hsize,trim={1.4cm .9cm .8cm .5cm},clip]{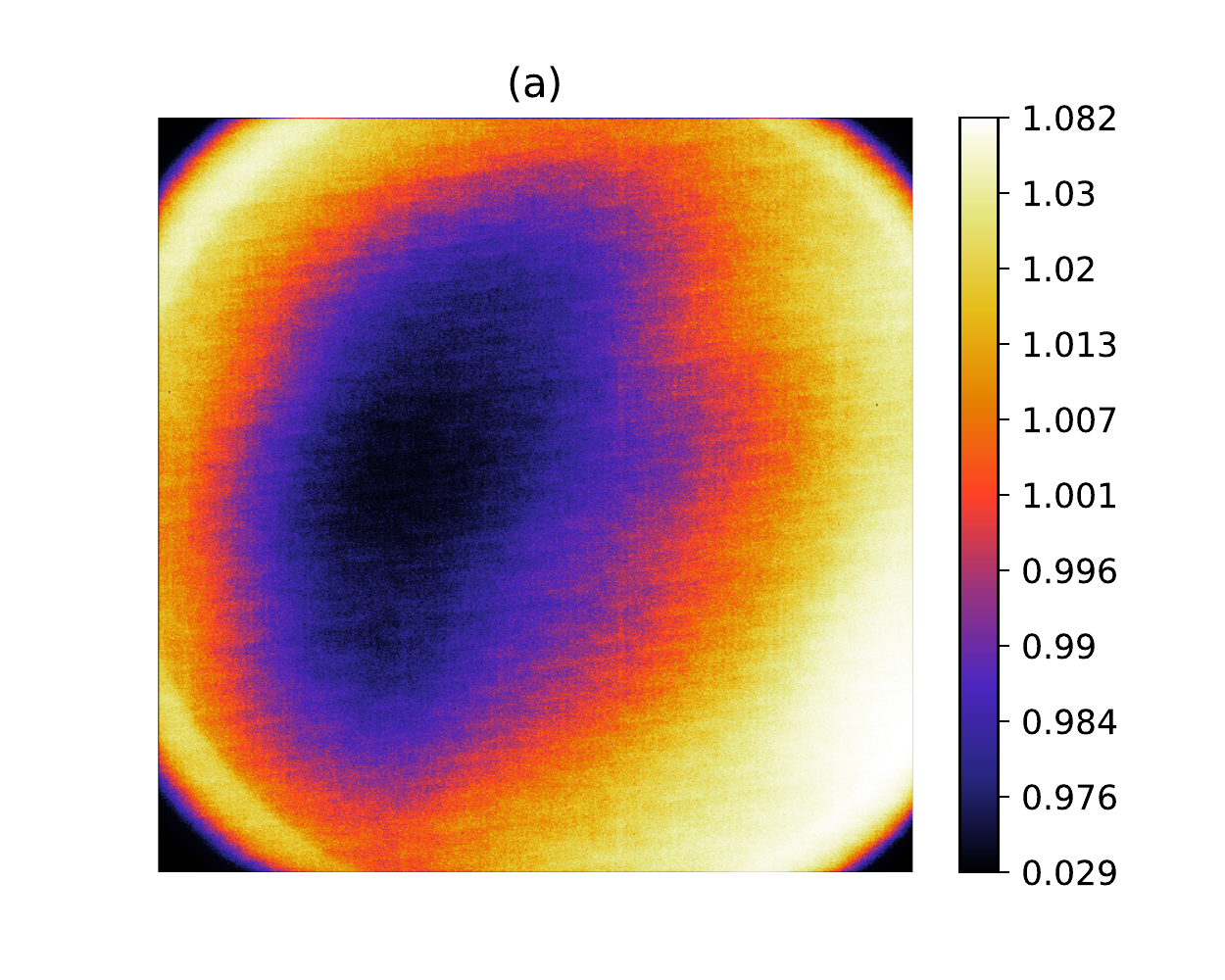}
				\includegraphics[width=.65\hsize,trim={1.4cm .9cm .8cm .5cm},clip]{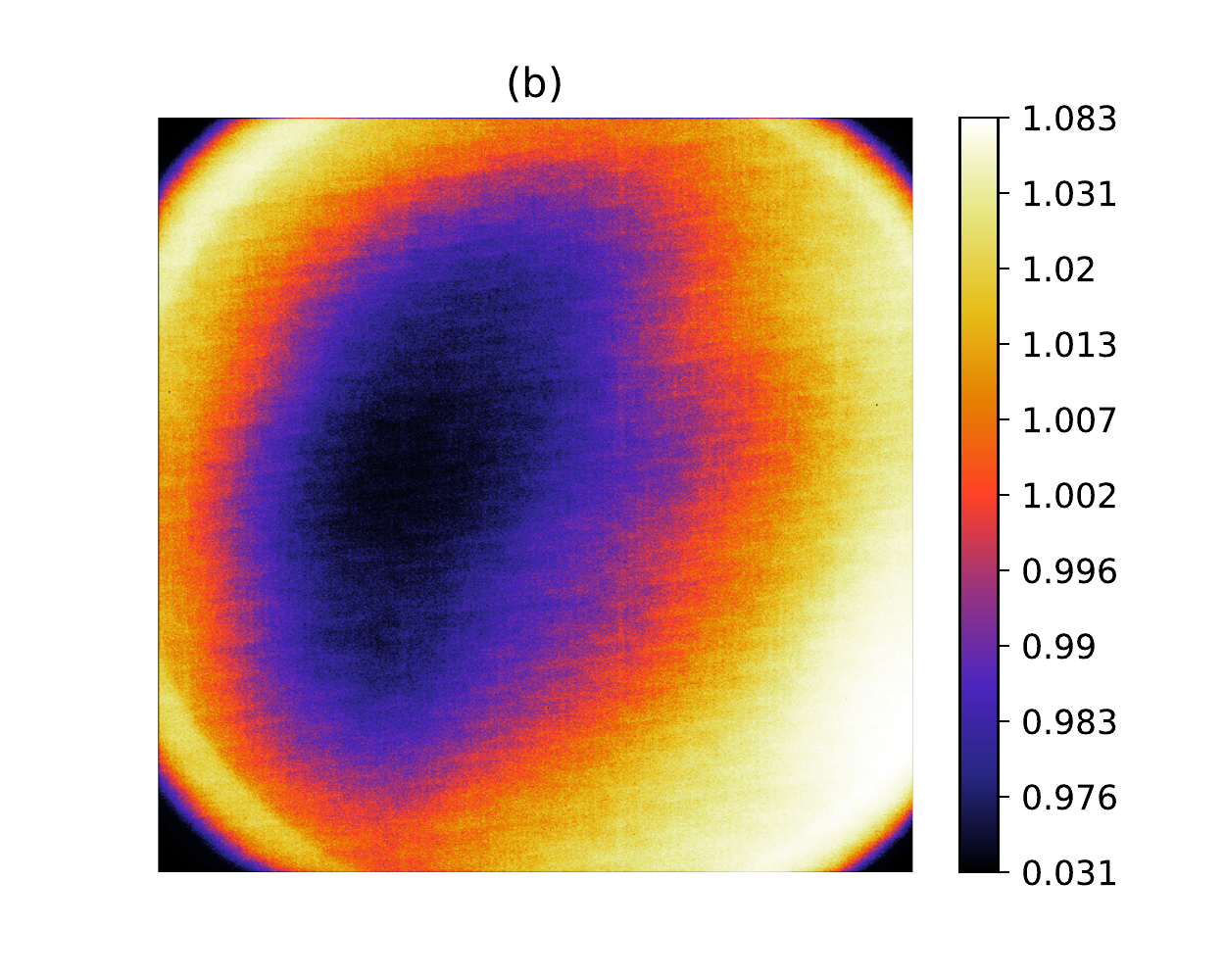}
				\includegraphics[width=.65\hsize,trim={1.4cm .9cm .8cm .5cm},clip]{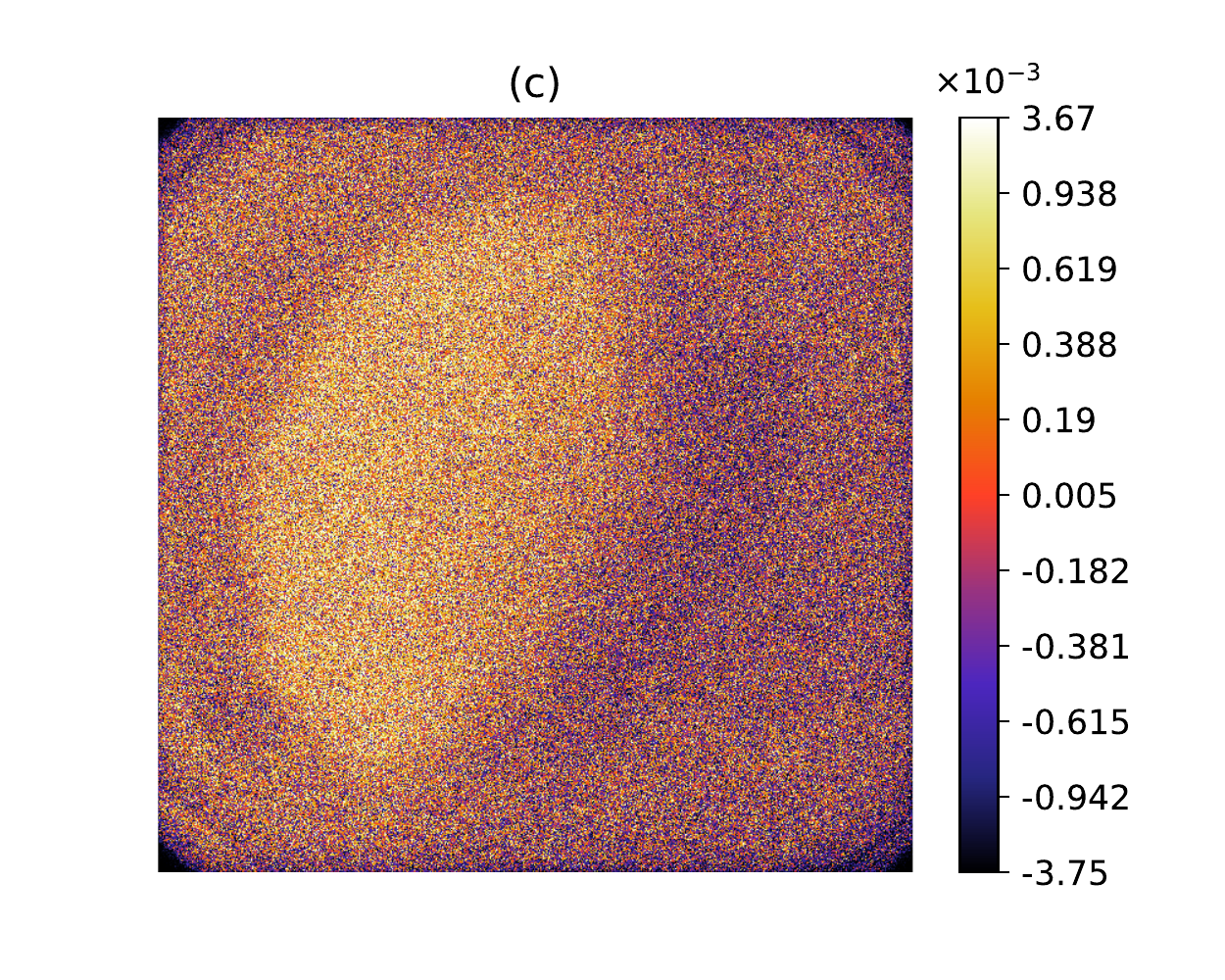}
				\caption{Results of the flat-field synthesis. (\emph{a})~Target flat field measured with the Tungsten lamp. (\emph{b})~Synthesised flat field. (\emph{c})~Residual image with a dispersion of~0.073\%\,rms, slightly greater than the noise-limited precision~(0.058\%). The image colour scales are expressed in normalised number of photo-electrons. They are not linear and have been optimised to highlight small variations using histogram equalisation.}
				\label{fig:ffc_res}
			\end{figure}

			Our ability to establish an overall flat-field response from chunks of flat fields is essential to accurately correct it for a large range of stellar types. To demonstrate we could produce "\`a la carte" flat-field synthesis, we uniformly illuminated the CCD with a Tungsten lamp (2500\,K) and we measured its corresponding flat field. To account for the transformation of the source emission spectrum into photo-electrons by the payload (optical telescope and CCD), we computed the spectral distributions of the Tungsten and calibration flat fields in \emph{electrons}. We then established the best linear combination of our series of filter spectral distributions (discarding the U broad-band filter) to reproduce the observed tungsten spectral shape (see Fig.~\ref{fig:ffc_sp}). We used these coefficients to compute the linear sum of the filter flat fields and normalised it. We compared our synthetic flat field with the measured one. We found a dispersion of residuals of $7\,10^{-4}$\,rms (see Fig.~\ref{fig:ffc_res}) corresponding to the average rms precision per pixel for each flat field. We also noticed a spatial structure in the residuals with an amplitude of about~0.06\%, which shape seemed to indicate a slight under-correction in the infrared.
			
			The light source used in this validation test reproduces the spectrum of a very cool star. For hotter stars, the fine sampling available at short wavelengths guarantees an even better synthetic flat field.

\section{Photometric performances} \label{sec:cal_phot}
	
	This section reports on the analyses and results from long-term series of measurements obtained during the calibration campaign with the purpose of assessing the photometric performances of CHEOPS.
	
	\subsection{Measurement} \label{ssec:cal_meas}
		
		During 27.35 hours, we collected an uninterrupted series of data, with an exposure time of 3\,s, simulating an observation equivalent to the duration of 16 CHEOPS orbits. During this long sequence, the payload was operated in nominal science conditions (R-channel read-out, CCD at \text{-40\degr C}, nominal bias voltages and read-out frequency of 230\,kHz). The front-end electronics was stabilised at \text{-5\degr C}, \text{5\degr C} warmer than its nominal value, without consequences on operation performances. The detector of the instrument was uniformly illuminated using the extended-source mode of the FPM and the light source was operated in open-loop mode. To avoid photometric variations due to intensity variation of strong emission lines, we used a top-hat filter to select a continuous wavelength range, between 500\,nm and 800\,nm (see Fig.~\ref{fig:ffc_sp}), where the source spectrum is free from such features. A total of 12243~images were recorded and analysed.

	\subsection{Data processing} \label{ssec:cal_proc}
		
		\subsubsection{Raw light curve} \label{sssec:cal_raw_lc}
			
\begin{figure*}
				\centering
				\includegraphics[width=.75\hsize,trim={1.2cm 1cm 2.7cm 2.4cm},clip]{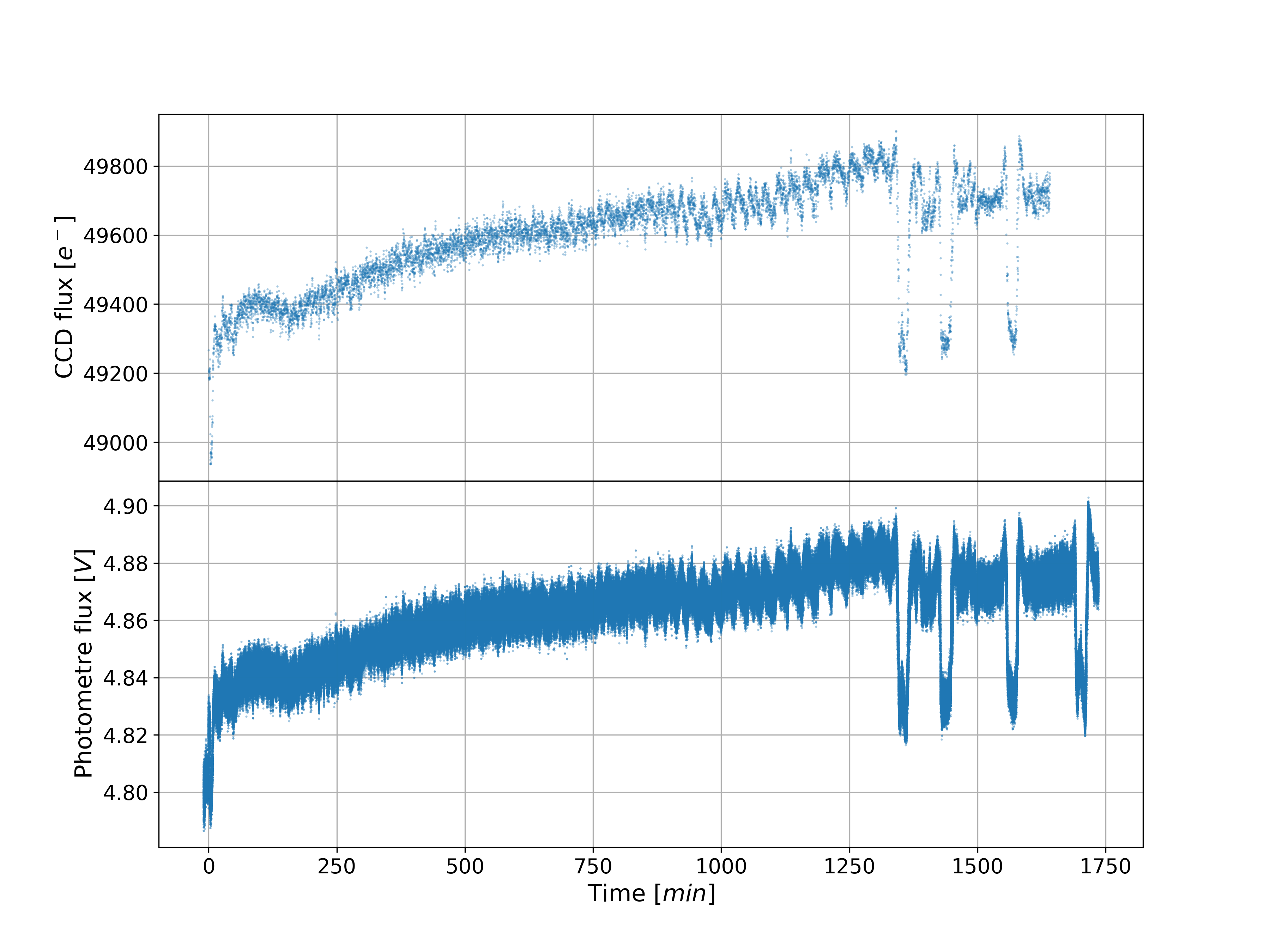}
				\caption{\emph{Top}. Raw light curve extracted from the images with a sampling time of 8.04\,s. \emph{Bottom}. Light source variations measured by the feedback photometer at a sampling frequency of about 12\,Hz.}
				\label{fig:pp_raw_lc}
			\end{figure*}

			Each frame is corrected for its bias level, converted to electrons and corrected from non-linearity effects and dark current. We $\sigma$\text{-clipped} each image to correct cosmic hits and bad pixels (two iterations with a 4-$\sigma$ threshold). Finally, we computed the mean flux on the whole image and used it as our photometry measurement. The photometric sequence (or light curve equivalent) is visible on Fig.~\ref{fig:pp_raw_lc}. The effects of open-loop operations, without feedback regulation of the lamp, are obvious on this figure.
			
		\subsubsection{Correcting the data from the light source variability} \label{sssec:cal_source_corr}
			
			The constant monitoring of the source flux by the photometer provided a light curve sampled at about 12 Hertz (see bottom panel of Fig.~\ref{fig:pp_raw_lc}). We used this high-frequency data to correct the CCD light curve from the source variability.
			
			During the measurements, the two data sets were not synchronised in time. We estimated the timing mismatch by cross-correlation assuming they have similar pattern and we measured a time delay of 13.24\,s. We shifted and resampled the photometer data accordingly to match the CCD acquisition rate with duration corresponding to CCD exposure time (3\,s) and sampling time of (8.04\,s). Finally, we normalised the photometer data by its average voltage by which we divided the CCD curve. The corrected curve is display on top panel of Fig.~\ref{fig:pp_raw_temp}. Variable features at~1\% level visible on the light curve before correction have been removed. The corrected light curve shows peak-to-valley amplitude variability of less than 200\,ppm.

		\subsubsection{Residual temperature correlation} \label{sssec:cal_temp_corr}
			
\begin{figure*}
				\centering
				\includegraphics[width=.69\hsize,trim={.4cm 1.8cm 2.5cm 3cm},clip]{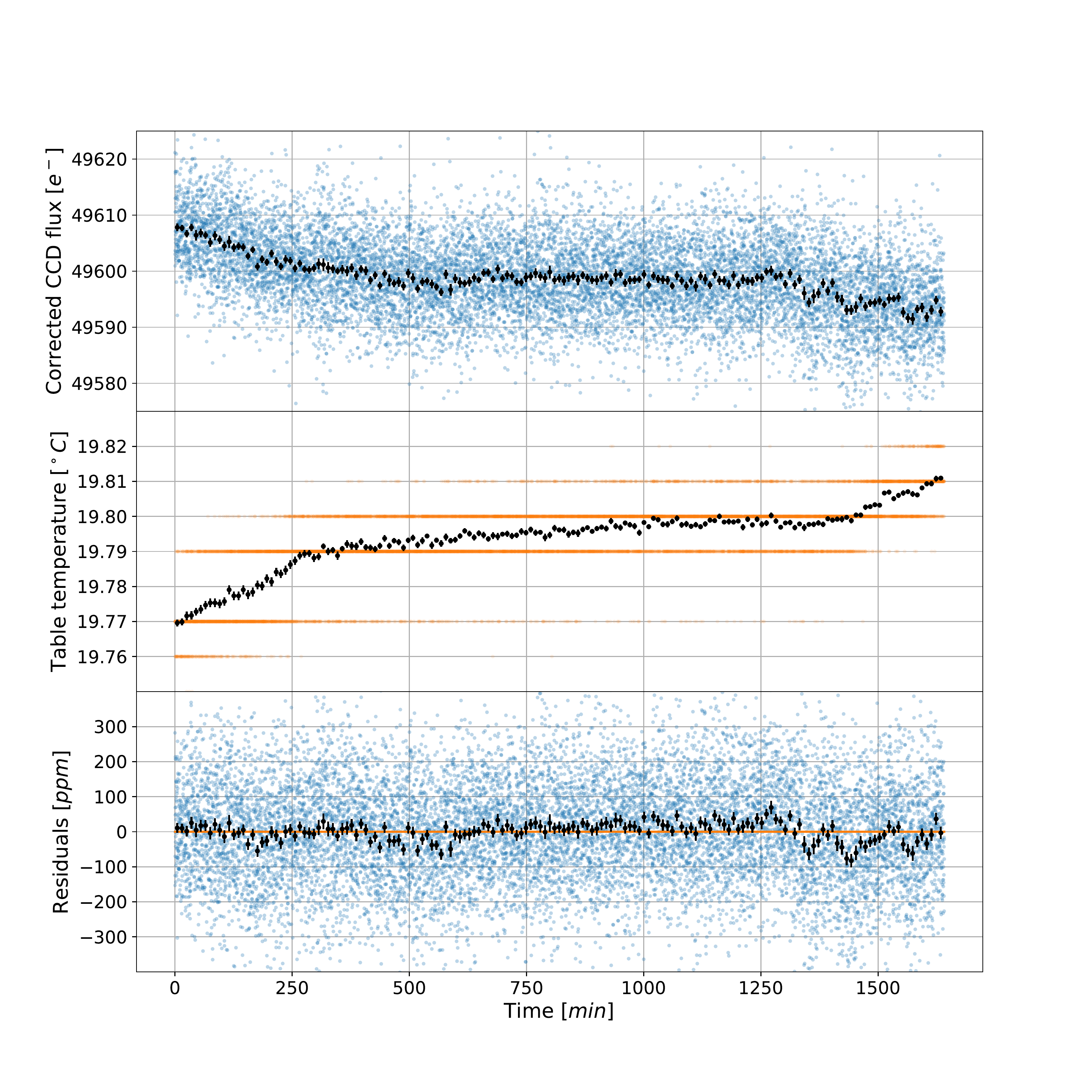}
				\caption{\emph{Top}. Light curve corrected for the source variability. \emph{Middle}. Temperature of the optical table. \emph{Bottom}. Residuals after correction of temperature correlation. \emph{All}. The \emph{black} points represent the 10-minute data binned.}
				\label{fig:pp_raw_temp}
			\end{figure*}

			The thermal sensitivity of the lamp regulation feedback (mostly due to the feedback fibre) affects the values measured by the photometer such as they do not accurately reflect the flux received by the CCD. We expect a correlation with the laboratory temperature change. A PT100 thermal sensor is attached on the optical table and records its temperature for each CCD exposure with a resolution and precision of 0.01\degr\,C. When one bins over 10 minutes the data of the temperature sensor, a gentle drift with time is visible and obviously anti-correlated with the corrected light curve (see the two top panels of Fig.~\ref{fig:pp_raw_temp}).
			
\begin{figure}
				\centering
				\includegraphics[width=\hsize,trim={1.8cm .5cm 3.3cm 2cm},clip]{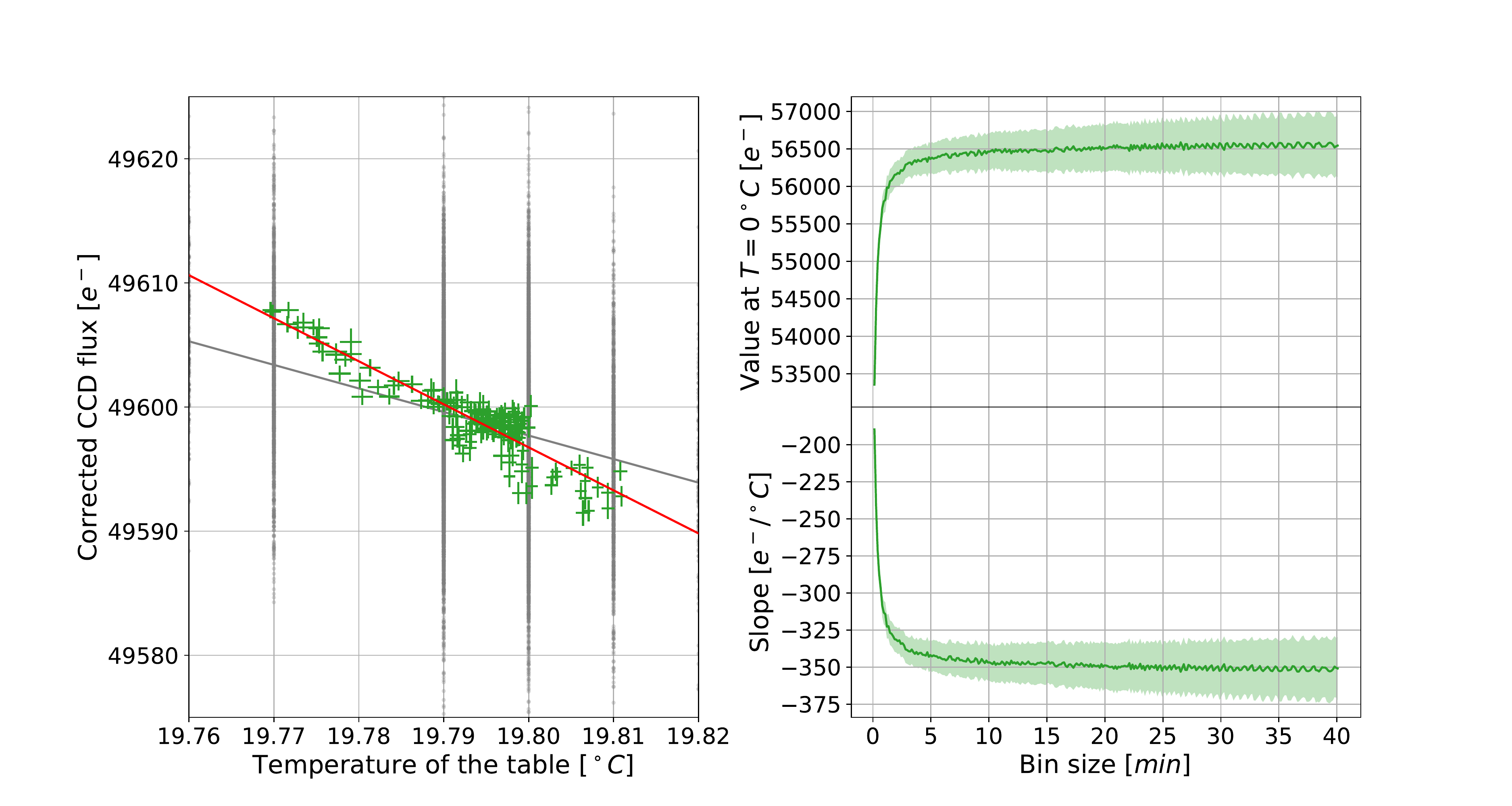}
				\caption{\emph{Left}. Temperature-flux scatter plot (\emph{grey}), with the 10-minute binned data overplotted in \emph{green}. The respective linear fits are the \emph{grey} and \emph{red} solid lines. \emph{Right}. Linear-fit parameters (slope and value at $T=0$\degr C) versus time-length considered for binning.}
				\label{fig:pp_ttable}
			\end{figure}

			No time delay between the light and temperature curves was detected. With the goal to decorrelate the light curve variations using the measurement of the optical table temperature, we calculated the coefficient of linearity between the two data sequences. We noticed that this coefficient was depending on the time sampling resolution of the thermal sensor considered as expected by its limited resolution. We found that, when the temperature measurement is averaged over 10-minute durations, the error due to the limited resolution of the PT100 cancelled out and the linear coefficient converges to $-347\,\text{e}^-$/\degr C (see Fig.~\ref{fig:pp_ttable}). To decorrelate the flux from the measured temperature variations, we used this correlation parameter and correct each photometric data point with its instantaneous temperature measurement. The corrected and de-trended light curve is shown on Fig.~\ref{fig:pp_raw_temp}.

	\subsection{Photometric precision} \label{ssec:cal_perf}
		
\begin{figure*}
			\centering
			\includegraphics[width=.75\hsize,trim={2.3cm .5cm 3cm 1.7cm},clip]{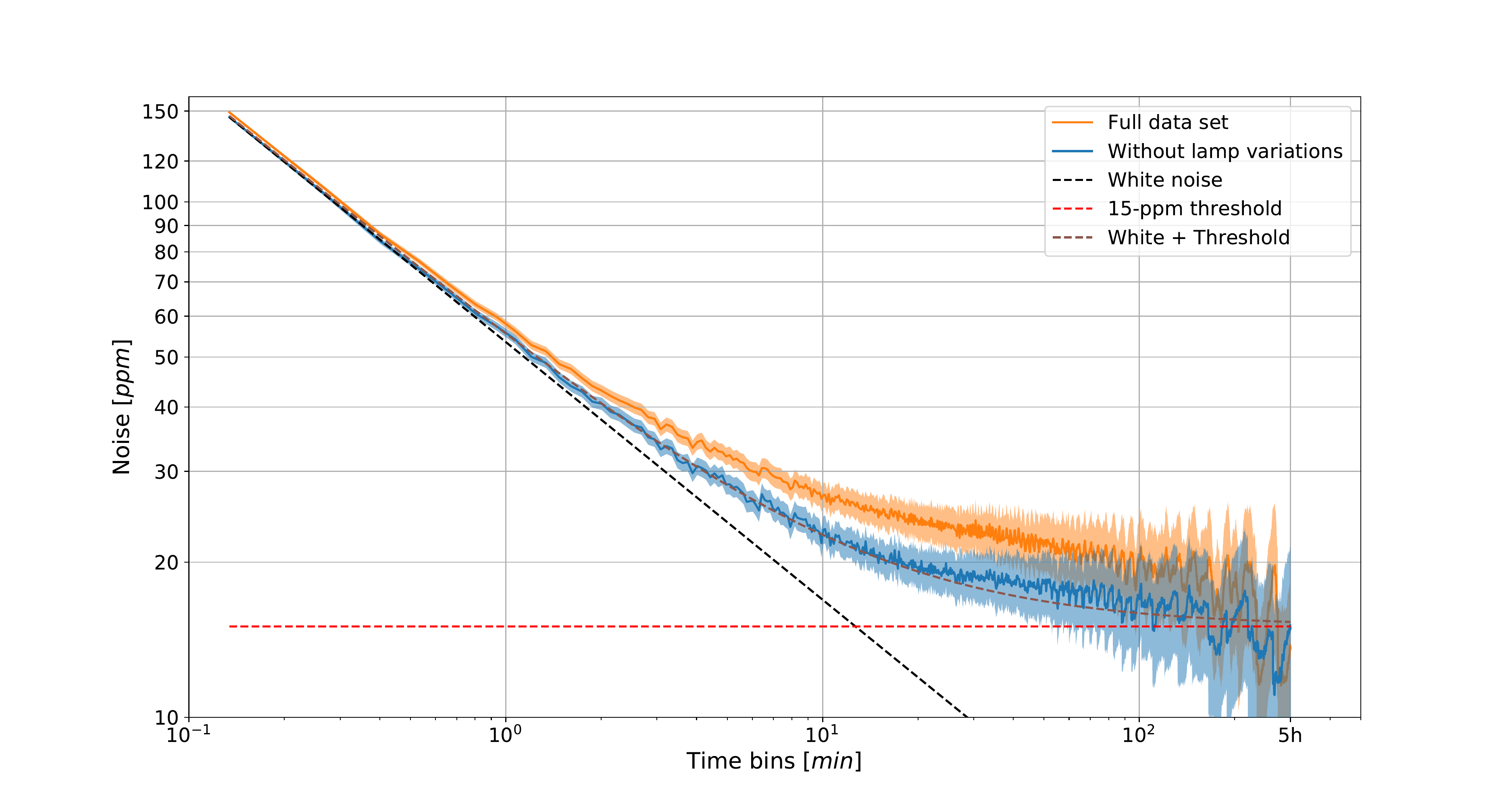}
			\caption{Noise curve of corrected and de-trended light curve (\emph{orange}) with the estimated errors represented by the shaded areas. The \emph{blue} curve shows the result after discarding the points taken during abrupt variations of the lamp ($t$\;>\,1320\,min). The noise values are not computed for time bins greater than 5~hours to ensure at least four points in the binned data. The \emph{brown} dashed curve is the quadratic sum of the white noise slope (oblique \emph{black} dashed line) and an arbitrary threshold at 15\,ppm (horizontal \emph{red} dashed line).}
			\label{fig:pp_nc_all}
		\end{figure*}

		The Allan variance analysis of the corrected and de-trended light curve is displayed on Fig.~\ref{fig:pp_nc_all}. The visible deviation from the expected white noise regime is suggesting a noise floor and an additional correlated noise structure is present in the data. One can notice that if one discards data points taken during the abrupt variations of the lamp, corresponding to measurement taken at~$t$~>\;1320\,min, the variation of the data Allan variance may be modelled using a white noise model and a noise floor of 15\,ppm. It is realistic to believe that during these extremes variations of the lamp, second-order effects may not be properly accounted and corrected by a simple flux normalisation. Unfortunately, strict deadlines to deliver the payload on time did not allow us to repeat the measurements to confirm this assumption.
		
		So far, in our analysis, the whole image area of the CCD have been considered. Practically, to be compared with a photometric extraction with a circular aperture of 33~pixels in radius as it will be done during target observations \citep{hoyer_cheops-drp}, one needs to consider the case of a limited number of pixels. In that case, the signal-to-noise ratio of a light curve can be computed according to the number of pixels considered with the following equation: $\sigma_{tot}/n_{e^-}=\left(n_{px}\times \text{RON}^2+\sigma_{e^-}^2\right)^{1/2}/n_{e^-}$, with $n_{px}$, $n_{e^-}$ and $\sigma_{e^-}$ being respectively the number of pixels, the number of photo-electrons and the photo-electron shot noise. In addition, the number of photo-electrons in the photometric aperture is limited by the fact that the flux is adjusted such as the value of strongest peak of the PSF in a pixel never reaches the saturation of the detector. According to the shape of the CHEOPS PSF, we know that the value of the flux contained in a single pixel represents at maximum 2\% of the integrated flux in the PSF. With this constraint, a PSF peak value equal to 70\% of the CCD dynamical range (typical to prevent saturation) gives a total number of electrons in the whole circular aperture (33~pixels in radius) that corresponds to a relative noise of 522\,ppm. Using the same equation in the case of uniformly illuminated images and a flux of $49600\,\text{e}^-$/pixel, we find 74~pixels would generate similar signal-to-noise level.
		
\begin{figure}
			\centering
			\includegraphics[width=\hsize,trim={1.2cm .7cm 2.2cm 2.1cm},clip]{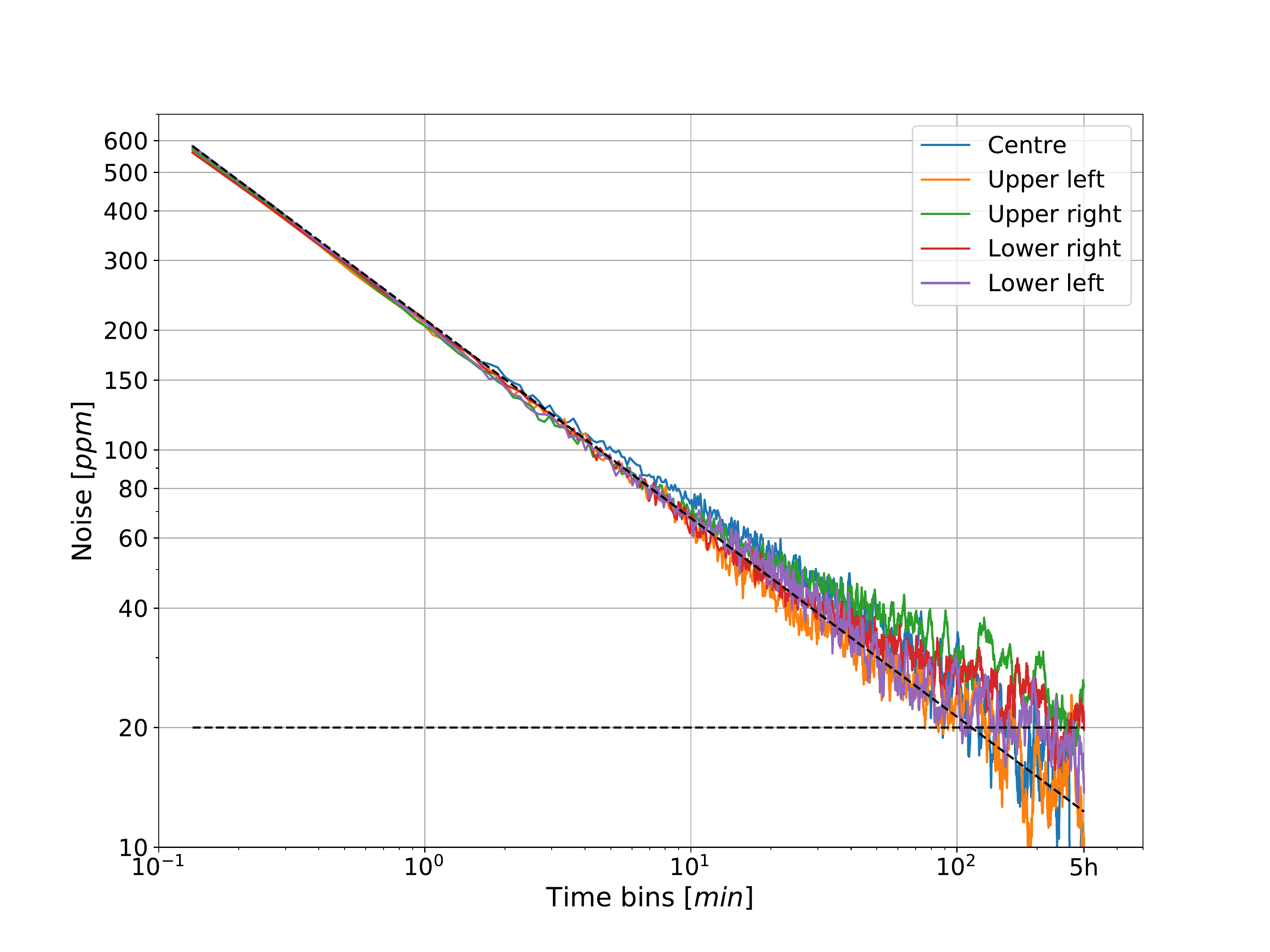}
			\caption{Noise curves of the flux extracted from five 8$\times$8-pixels windows located at different positions on the detector (as indicated by the legend). The oblique \emph{black} dashed line indicates the theoretical white noise slope while the horizontal one illustrates the 20-ppm level.}
			\label{fig:pp_nc_sub}
		\end{figure}

		We then repeated our photometric analysis, extracting only 8$\times$8-pixel regions of our CCD. This would correspond to a PSF with a maximum value of $75\,\text{ke}^-$ (62\%~of the dynamical range). Five light curves were computed selecting different areas of the CCD. The overall analysis of the data is identical to the one applied previously and described in sections above. Fig.~\ref{fig:pp_nc_sub} displays the result of the Allan analysis of the first 22~hours of the five light curves. In all cases, the noise follows the white noise regime and reach 20\,ppm in 5\,hours, in accordance with mission design requirements.
		
\section{Photometric performances from simulated observations} \label{sec:sim}
	
	In this chapter, we investigate the expected in-flight performances by conducting and end-to-end simulation based on a realistic payload simulator (\texttt{CHEOPSim}) and results from the calibration campaign. Two sequences of observations were simulated. Each one can be related to a core science requirement of the mission. In the next section, we describe how the data have been produced, in the following section the data processing is detailed and finally the results will be presented in terms of residual noise and precision on determination of the planet-to-star radii ratio.
	
	\subsection{Simulated observations} \label{ssec:sim_data}
		
		\subsubsection{The \texttt{CHEOPSim} simulator} \label{sssec:cheopsim}
			
			\texttt{CHEOPSim} is a software developed for the CHEOPS mission that simulates the scientific data produced by the CHEOPS payload \citep{futyan_cheopsim}. It produces series of CCD images in a format similar to the one that will be available for science analysis on the ground. It includes the following features: the modelling of the stars (limb darkening, granulation, activity), the modelling of the transits, the modelling of the background (stray light, zodiacal light, cosmic rays), the modelling of the satellite (orbit, pointing jitters, field of view rotation, South Atlantic Anomaly, Earth occultation), the modelling of the telescope optics (optical transmission, PSF) and the modelling of the detector (bias offset, read-out noise, dark current, gain, non-linearity, flat field, quantum efficiency, bad pixels, saturation, frame transfer, smearing trails, charge-transfer inefficiency). Time series produced are discontinued with interruptions corresponding to South Atlantic Anomaly crossings (when data are not down-linked to ground in order to save telemetry) and when the target line of sight is occulted (by Earth) or too close to the stray-light limit (minimum angle separation to the sunlit limb of the Earth).
			
			All features included in the modelling of the detector are based on performance measurements done during payload calibration. \texttt{CHEOPSim} uses PSF measured on the ground during the calibration campaign (see Fig.~\ref{fig:sim_image}). Note that PSF shape is likely to be slightly different in space due to the absence of gravity. (For more details on \texttt{CHEOPSim}, see \citealt{futyan_cheopsim}.)

		\subsubsection{Description of the data sets} \label{sssec:sim_descr}
			
			We simulated data sets of two series of observations representative of the mission requirements introduced in section~\ref{sec:intro}, with the purpose of validating the capability of CHEOPS in fulfilling its scientific goals. The first series of data simulates a sequence of two consecutive transit observations of an Earth-size planet orbiting a Sun-like star (G2V) with a period of 50~days. The second series simulates a single transit of a Neptune-size planet orbiting a K5V~star with a period of 13~days. See Table~\ref{tab:sim} for more details about simulation parameters. In both cases, the planetary orbits are circular and the transit geometry has an impact parameter $b=0$. Stellar photosphere effects, like spots or granulation, were not considered but the three-dimension nature of the atmospheres has been taken into account (limb-darkening effect).
			
			\begin{table}
				\caption{Parameters of the simulated data sets.}
				\label{tab:sim}
				\centering
				\begin{tabular}{l l|c@{\hspace{1.8em}}c}
					\multicolumn{2}{c}{Parameters} & Case 1 & Case 2 \\[.4ex]
					\hline\hline
					\multirow{6}{*}{Star} & Spectral type & G2V & K5V\rule{0pt}{2.6ex} \\
					& T$_{\text{eff}}$ & $5770\ K$ & $4450\ K$ \\
					& V-magnitude & 9 & 12 \\
					& Radius & $1.01\ R_\odot$ & $0.709\ R_\odot$ \\
					& Mass & $1.02\ M_\odot$ & $0.72\ M_\odot$ \\
					& Activity & none & none \\[.4ex]
					\hline
					\multirow{4}{*}{Planet} & Radius & $1\ R_\oplus$ & $1\ R_\text{\Neptune}$\rule{0pt}{2.6ex} \\
					& Orbit & circular & circular \\
					& Period & 50 days & 13 days \\
					& Inclination & 90\degr & 90\degr \\[.4ex]
					\hline
					\multirow{4}{*}{Background} & Star field & \multicolumn{2}{c}{BD-082823}\rule{0pt}{2.6ex} \\
					& Stray light & \multicolumn{2}{c}{uniform} \\
					& Zodiacal light & \multicolumn{2}{c}{uniform} \\
					& Cosmic rays & \multicolumn{2}{c}{none} \\[.4ex]
					\hline
					\multirow{3}{*}{Observation} & \# of transits & 2 & 1\rule{0pt}{2.6ex} \\
					& Duration & 2 $\times$ 20h & 10h \\
					& Cadence & 1 min & 1 min \\[.4ex]
					\hline
					\multirow{2}{*}{Interruptions} & Stray light & \multicolumn{2}{c}{when > 0.6 ph/px/s}\rule{0pt}{2.6ex} \\
					& SAA\tablefootmark{*} & \multicolumn{2}{c}{when crossing area} \\[.4ex]
					\hline
				\end{tabular}
				\tablefoot{\tablefoottext{*}{South Atlantic Anomaly.}}
			\end{table}
			
\begin{figure}
				\centering
				\includegraphics[width=\hsize,trim={2.6cm 2cm 3.5cm 1.2cm},clip]{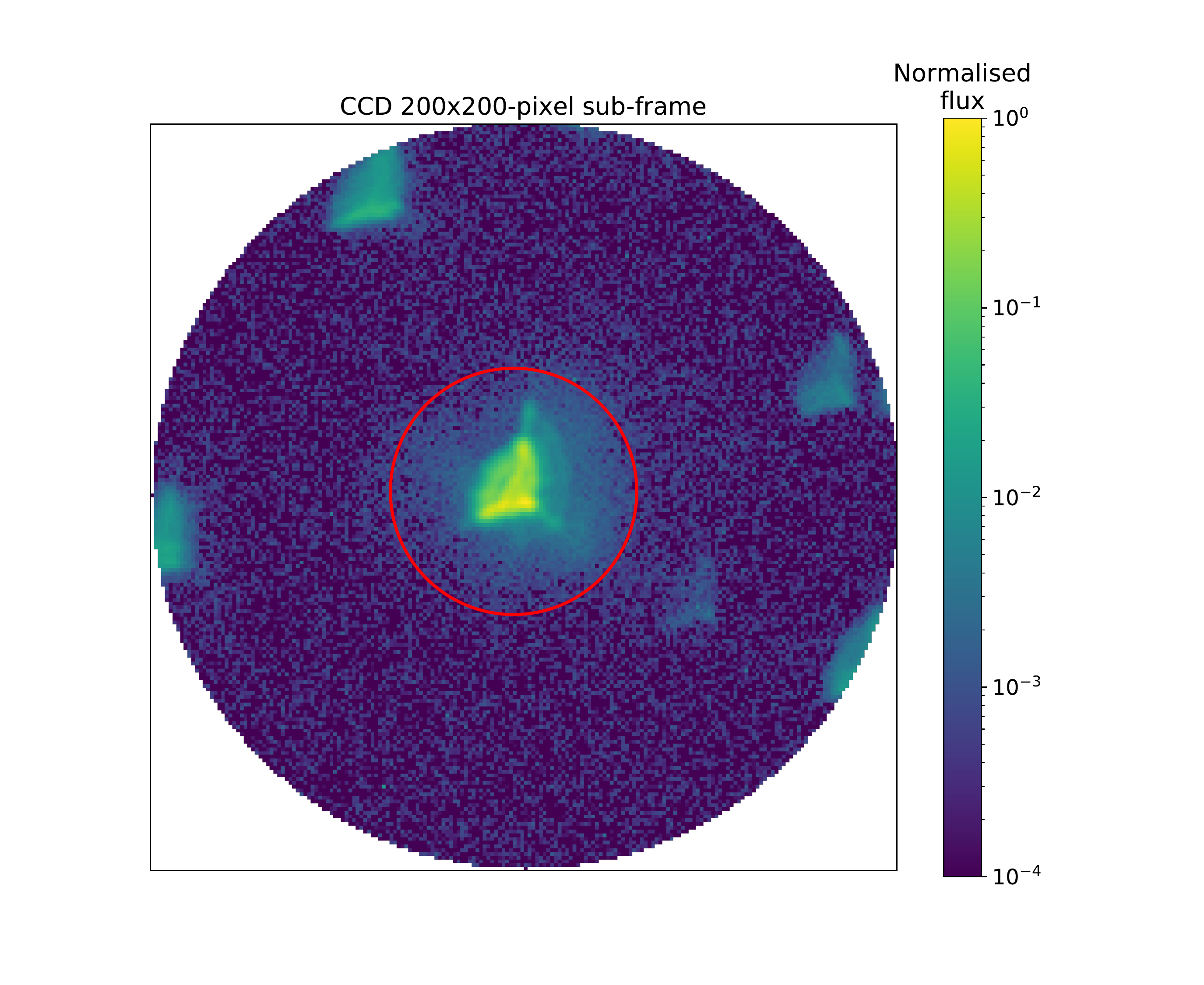}
				\caption{Simulation image computed by \texttt{CHEOPSim}. The central target star has a \text{V-band} magnitude of~12, while the background field is the one of \text{BD-082823}. The colour scale is the flux corrected for all instrumental and background signals, and normalised to the highest peak of the central PSF. The \emph{red} circle represents a typical photometric aperture with a radius of 33~pixels.}
				\label{fig:sim_image}
			\end{figure}

			The star field used in the background is the actual real stellar background one would found when pointing the star \text{BD-082823}, corresponding to a galactic longitude and latitude of respectively 248.4966\degr\ and~+34.7560\degr. It represents a field crowding configuration typical of most of CHEOPS observations (see Fig.~\ref{fig:sim_image}).
			
			The focus of this work is to evaluate the impact of the combination of satellite jitter, detector features, PSF shape and stellar background on scientific results. To simplify the analysis, the cosmic rays and smearing trails were not included in our simulation. This assumption is actually equivalent to consider that the correction is efficient enough for the cosmic rays to have a marginal impact on the data and the smearing model is good enough to correct this effect in the case of very short exposure times (not the case considered in our simulations).

	\subsection{Data analysis} \label{ssec:sim_red}
		
		We developed a photometric extraction package to analyse \texttt{CHEOPSim} data. Our software computes photometry of the target from the raw images generated by the simulator. When executed on the whole series of data, it produces a light curve of the target star. This software is independent from the CHEOPS automatic data reduction pipeline (DRP) operated by the Science Operation Centre to reduce CHEOPS observations \citep{hoyer_cheops-drp}. We would like to point out that our software development served as well the purpose to provide us with a validation tool independent from the DRP during review phases.
		
		The electronic bias offset is estimated from the overscan regions and subtracted uniformly from the image. The image is converted from ADU to electrons with the gain and the non-linearity established during the calibration. Similarly to the bias level, the dark current is computed from active CCD pixels not exposed to the sky, and removed from the image. We synthesised the flat field corresponding to the spectrum of the target (see section~\ref{sssec:ff_syn}) and corrected the photo-response non-uniformity of the detector.
		
		To estimate the background level, we select CCD pixels located on a ring centred on the target star. The size of the ring inner radius is 50~pixels to exclude most of the flux from the target including its halo. The size of the outer radius is 90~pixels to keep the ring in the image when satellite pointing jitter~(4"\,rms) is considered. To compute a robust background value, we build the intensity histogram of pixels in the ring and we model it by a normal distribution ($\mathcal{N}\mathopen{}\left(\mu,\sigma^2\right)\mathclose{}$). The best-fit parameter $\mu$ corresponds to the mean background value.
		The PSF position is computed on the background-subtracted image by a classical iterative centroid measurement method. For each iteration, we weight the image with a two-dimension Gaussian function, and we calculate the new centre. This new position is used for the next iteration. The cycle stops when the centre remains within $10^{-4}$~pixel or after 30~iterations.
		As the background and the PSF centre depend on each other, our algorithm first assumes a PSF position at the image centre, estimates the background level, computes the PSF location and then repeats the last two steps from the new PSF centre.
		
		The photometric extraction is done by counting the number of electrons recorded on the CCD within a radius of 33~pixels centred on the target. The size of this radius is found to maximise the signal-to-noise ratio of the light curve. To maintain a constant number of pixels in the aperture, the intersection area of each pixel and the aperture disk is computed analytically and used to weight the pixel value located on the edge of the extraction zone.
		
		The errors on individual photometric data points of the light curve are computed from photon noise combined with the quadratic sum of all other measured uncertainties from calibration mostly dominated by gain and flat-field uncertainties (typically 0.6\% each). By comparison the precision on the bias level, dark current background are of the order of a few electrons.
		
		The light curve obtained from the photometric extraction is adjusted by a transit model implemented in \texttt{batman} \citep{kreidberg_batman}, with a quadratic limb-darkening model\footnote[1]{The quadratic limb-darkening model is given by the expression $I\mathopen{}\left(\mu\right)\mathclose{}/I_0=1-u_1(1-\mu)-u_2(1-\mu^2)$ where $I$ is the specific intensity on the stellar disk, $I_0$ its value at the centre of the disk, $u_1$ and $u_2$ are the coefficients, and $\mu=\sqrt{1-x^2}$ with $x$ being the normalised radial coordinate on the stellar disk.}. We use the \texttt{emcee} \citep{foreman_emcee} implementation of the Markov chain Monte Carlo (MCMC) algorithm to look for the best solution. \texttt{corner} \citep{foreman_corner} is used to visualise the multidimensional posterior distributions of the transit parameters from the MCMC outputs.
		
		The eccentricity is by definition fixed to zero. The out-of-transit flux~$F_0$ and the planet-to-star radii ratio~$R_p/R_*$ have uniform priors. The orbital period~$P$ and the mid-transit time~$t_0$ have Gaussian priors with typical uncertainties of respectively 12~hours and 60~hours for the first case (Earth size, bright star), and both 30~minutes for case~2 (Neptune size, faint star). To take into account the correlation between $t_0$ and $P$, the prior of $t_0$ is centred one orbital period before the observed transit. The prior of the semi-major axis~$a/R_*$ is also Gaussian-shaped assuming 5\%~uncertainty on the stellar mass and 10\%~on the stellar radius. The impact parameter $b$ is uniformly constrained in the interval ensuring the planetary transit ($b<1+R_p/R_*$). The uniform priors of the limb-darkening coefficients restrain their values to realistic ranges, as detailed by \cite{kipping_ld}.

	\subsection{Results} \label{ssec:sim_perf}
		
\begin{figure}
			\centering
			\includegraphics[width=\hsize,trim={.2cm 1.8cm 2.6cm 2.9cm},clip]{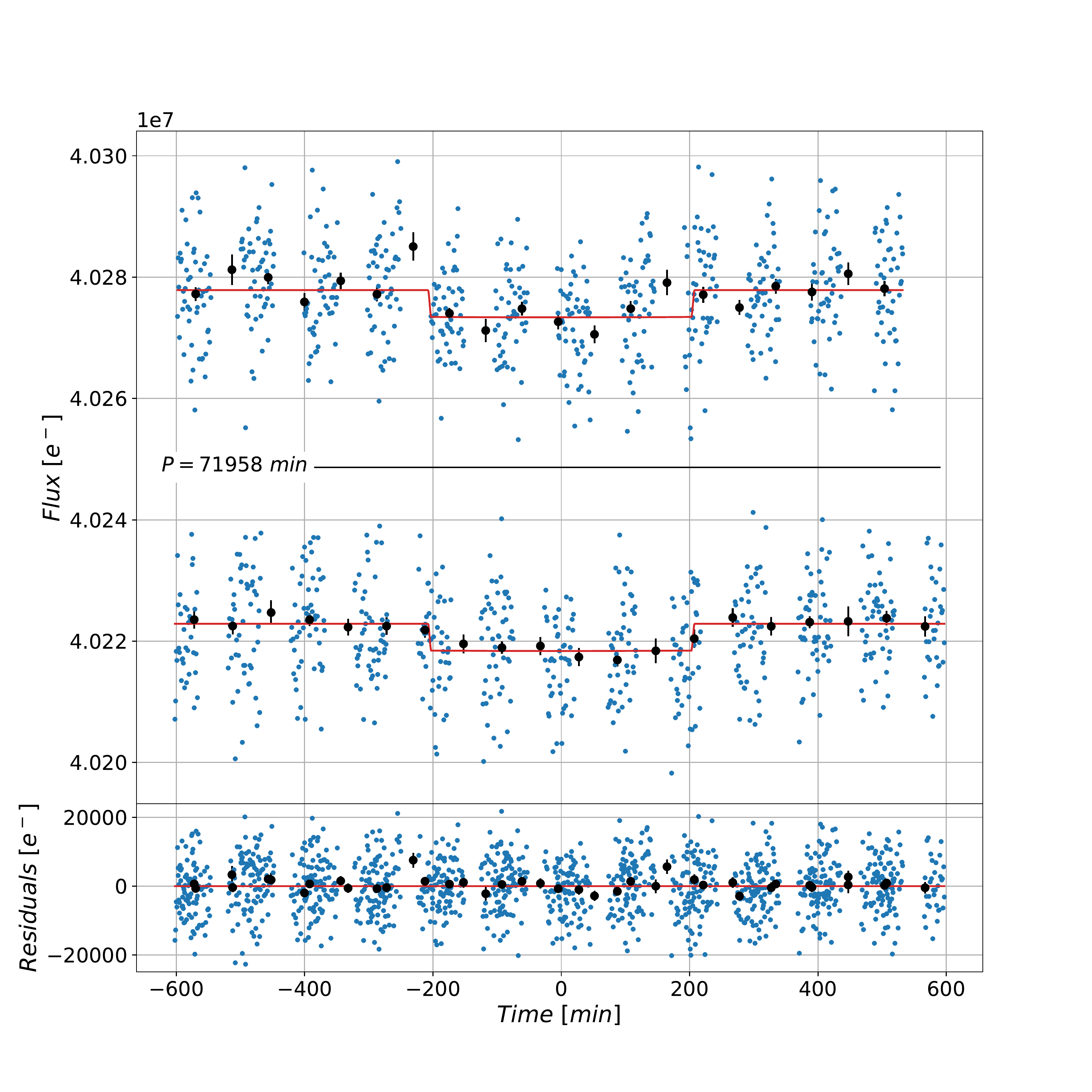}
			\caption{\emph{Top.} Light curves of the stellar flux extracted from the simulated images of two transits of an Earth-size planet in front of a Sun-like star (case 1). The two transits are phase-folded. The \emph{black} points are the 60-min binned data and the \emph{red} curve corresponds to the best-fit model. The regular data gaps visible in the series are related to pointing visibility limits of CHEOPS during its orbit. Time is expressed from mid-transit. \emph{Bottom.} Phase-folded residuals.}
			\label{fig:tlc_1}
		\end{figure}

\begin{figure}
			\centering
			\includegraphics[width=\hsize,trim={.1cm 1.8cm 2.6cm 3.1cm},clip]{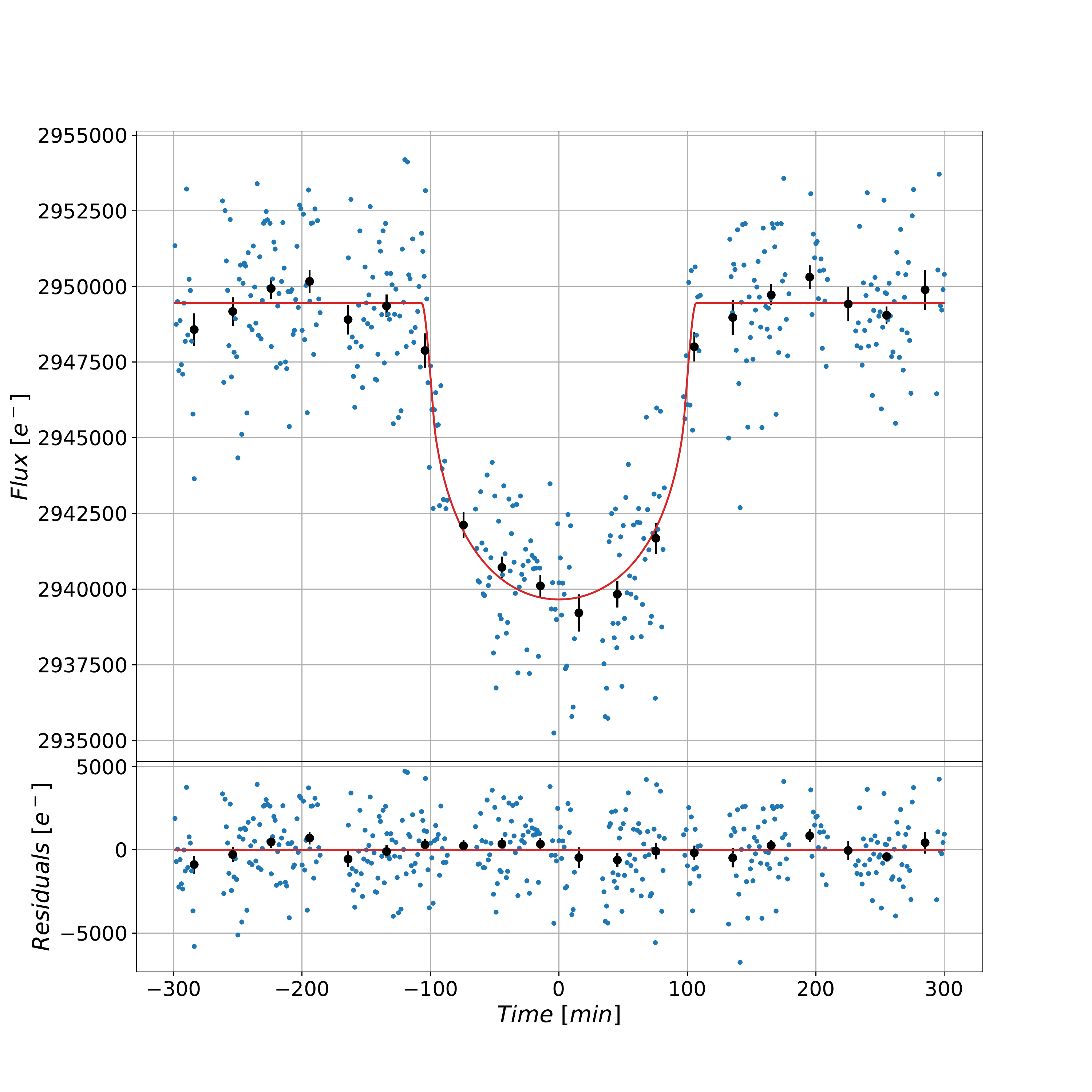}
			\caption{\emph{Top.} Light curve of the stellar flux extracted from the simulated images of a Neptune-size planet transiting a K5 dwarf star (case 2). The \emph{black} points represent 30-min binned data and the \emph{red} curve corresponds to the best-fit model. Time is expressed from mid-transit. \emph{Bottom.} Residuals of the best fit.}
			\label{fig:tlc_2}
		\end{figure}

		The phase-folded light curves and the results of our analyses are displayed on Fig.~\ref{fig:tlc_1} and~\ref{fig:tlc_2}. The gaps visible in the sequences are caused by regular observation interruptions each time the satellite crosses the South Atlantic Anomaly or when the target is occulted by the Earth. With the first case (Earth transits), we compute a rms of 10.2\,ppm when the residuals of the fit are averaged on the transit durations. Similarly a value of 51.7\,ppm is measured for the second case (Neptune transit). These values should be compared with the expected photon noise level (computed from the square root of the flux), respectively 8.3\,ppm (6 hours of data) and 43\,ppm (3~hours of data). The small differences are the contributions of other terms from the error budget (gain, jitter, flat field, etc.) considered by the end-to-end simulations.
		
\begin{figure*}
			\centering
			\includegraphics[width=.491\hsize,trim={0cm 0cm .6cm .2cm},clip]{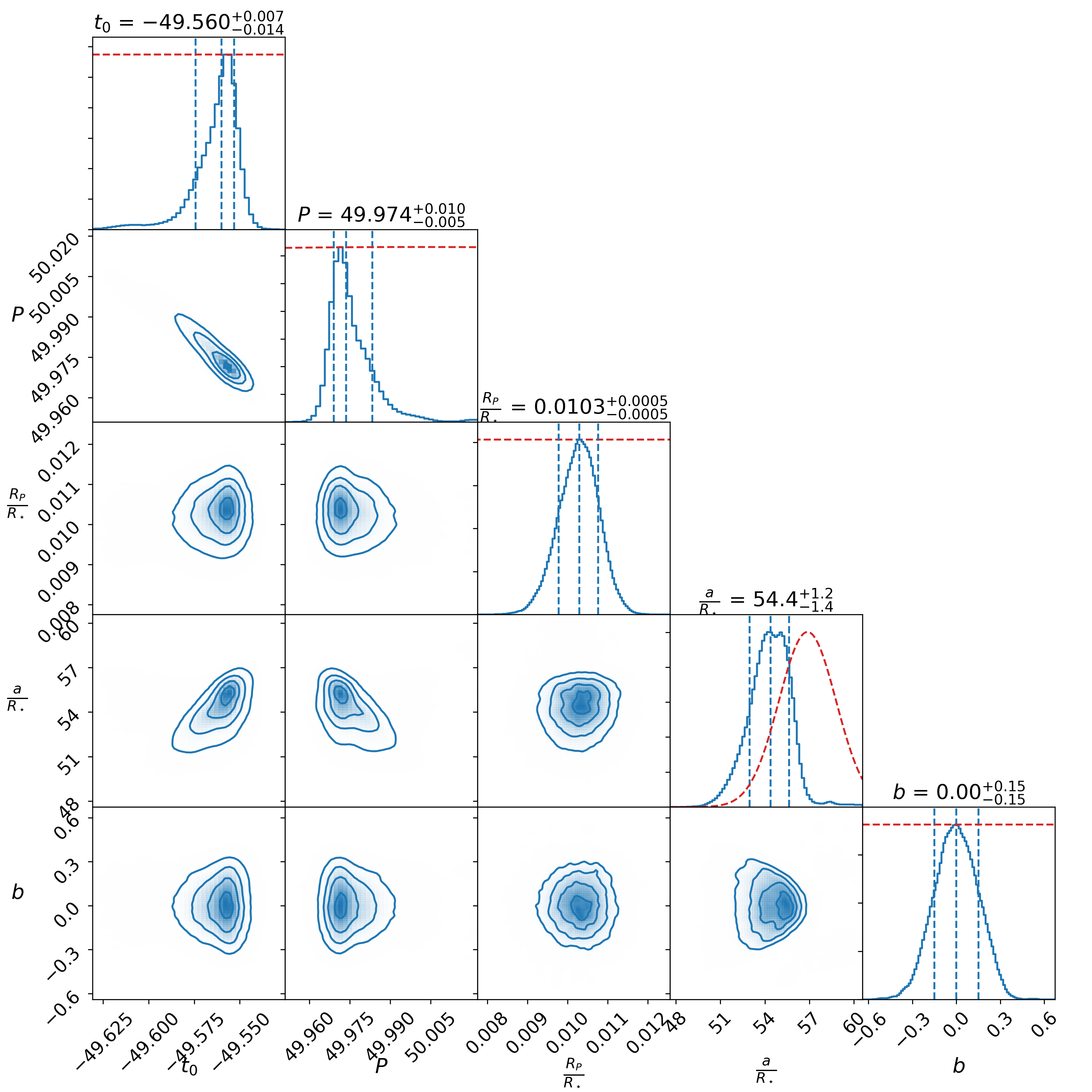}
			\space\space
			\includegraphics[width=.491\hsize,trim={0cm 0cm .6cm .2cm},clip]{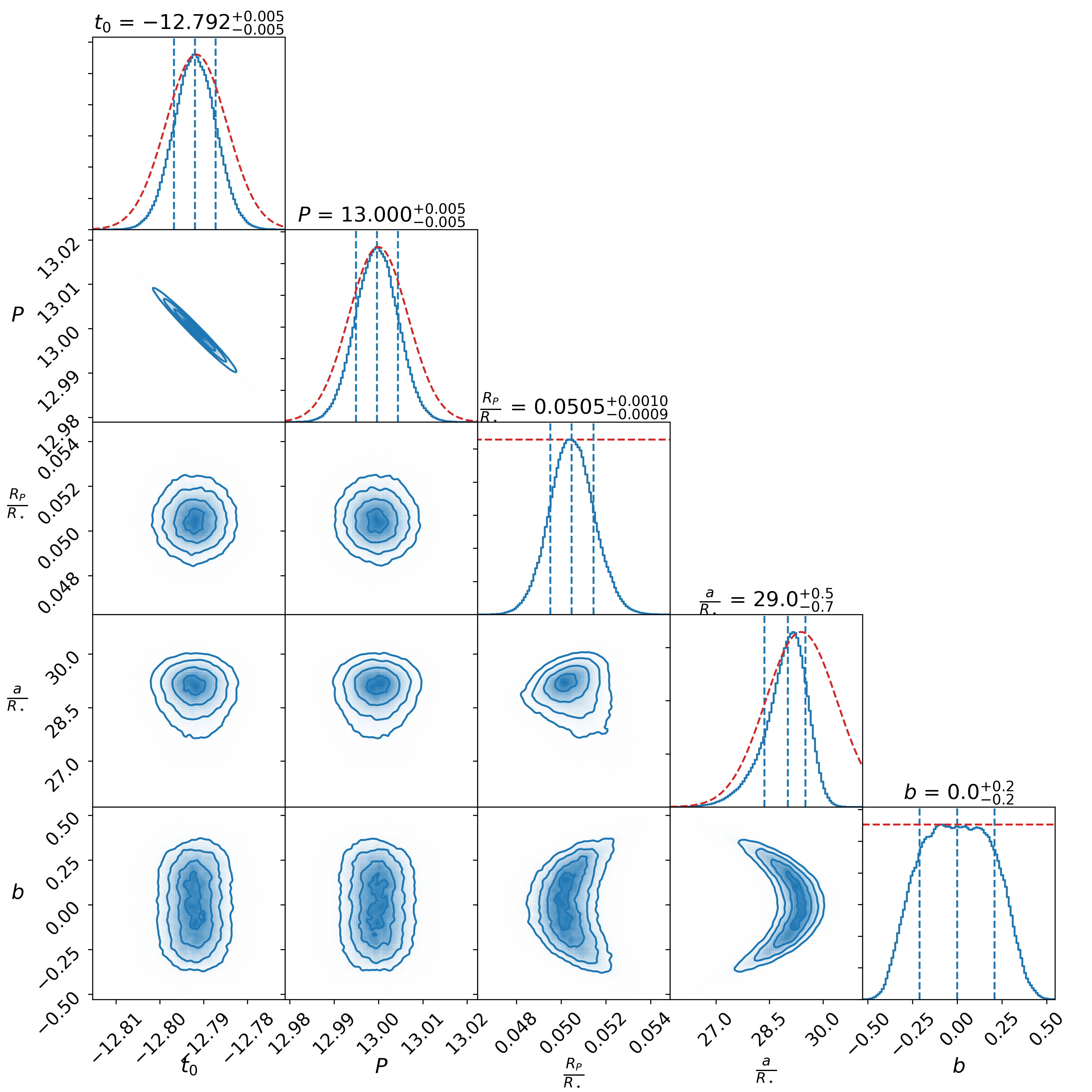}
			\caption{Corner plots of the posterior distributions of the transit parameters $t_0$, $P$, $R_p/R_*$, $a/R_*$ and $b$ obtained for case~1 (Earth -- \emph{left} graph) and case~2 (Neptune -- \emph{right} graph). The \emph{red} dashed lines represents the prior distributions. The \emph{blue} vertical dashed lines highlight the mean value and the 68\%~confidence interval of the marginalised distributions.}
			\label{fig:corner}
		\end{figure*}

		From corner plots of the MCMC outputs, we displayed five parameters showing interesting features: $t_0$, $P$, $R_p/R_*$, $a/R_*$ and~$b$ (see Fig.~\ref{fig:corner}). We note that correlations for the simulation of the Neptune planet (case~2) are stronger than for the Earth planet (case~1), possibly due to the higher SNR of the transit, especially the definition of the ingress and egress phases (see Fig.~\ref{fig:tlc_1} and~\ref{fig:tlc_2}). The strong correlation between $t_0$ and $P$ is explained by the fact that an overestimation of the orbital period tends to increase the duration between two transits and hence move the mid-transit time backwards in time. The effect of the impact parameter is visible by its correlation with the planetary radius and the semi-major axis. Indeed, when $b$ gets different from zero, the decrease of the duration and depth of the transit is balanced by a smaller $a/R_*$ (longer transit) and a larger $R_p/R_*$ (deeper transit).
		
		The corner plots display the marginalised distributions of the parameters along with a representation of the corresponding priors (red dashed lines on Fig.~\ref{fig:corner}). This allows us to see if the data constrain the parameters or if their values are only defined by the priors we initially set. Contrary to case~1, the absence of a second transit in case~2 prevents the parameters $t_0$ and $P$ to be constrained by the data and their prior and posterior distributions are almost identical. In both cases, the transit duration can be set by either $a/R_*$ or $b$, creating a degeneracy between these two parameters and explaining the prior-shaped posterior of the semi-major axis. One can note that the impact parameter is also constrained by the transit ingress and egress. A better coverage of these phases could potentially help lifting the degeneracy.
		
		The uncertainties on the parameter values are estimated for both cases by computing the 68\%-confidence intervals on the marginalised distributions (see Table~\ref{tab:perf}). In the case of the Earth-size planet, we obtain an uncertainty for the planet-to-star radii ratio of the order of~5\%, which corresponds to a precision on the transit depth of about~10\%, or a transit SNR of 10. With a transit depth of 100\,ppm, a SNR of 10 corresponds to 10\,ppm uncertainty on the transit depth. Our residual value on the light curve fit is consistent with this number (10.2\,ppm). For the Neptune-size planet, we obtain a precision of~2\% on $R_p/R_*$ (or 4\% on $(R_p/R_*)^2$), equivalent to a transit SNR of 25. The residuals of the fit (51.7\,ppm\,rms) and the transit depth of 2550\,ppm suggest a transit SNR of 49, twice higher, potentially leading to 1\% accuracy on $R_p/R_*$ . In that case, the transit SNR does not provide a good estimate of the expected precision on the radius. The correlation with the impact parameter and orbit interruptions tends to increase the uncertainty on $R_p/R_*$ measurement. A better prior knowledge of $b$ could potentially lead to the improvement of the results.
		
		\begin{table*}
			\caption[]{Transit parameters for the two cases, derived from the posterior distributions. The parameters $q_1$ and $q_2$ are computed from the coefficients $u_1$ and $u_2$ of the quadratic limb-darkening model\footnotemark[1], as detailed by \cite{kipping_ld}.}
			\label{tab:perf}
			\centering
			\begin{tabular}{l|c@{\hspace{1.8em}}c}
				Parameters & Case 1 & Case 2 \\[.6ex]
				\hline\hline
				Out-of transit flux [$\text{ke}^-$] & $40277.8\pm0.2$ & $2949.5\pm0.1$\rule{0pt}{3ex} \\[.6ex]
				Mid-transit time [days] & $-49.560\substack{+0.007\\-0.014}$ & $-12.792\pm0.005\ $\tablefootmark{$\dagger$} \\[.6ex]
				Orbital period [days] & $49.974\substack{+0.010\\-0.005}$ & $13.000\pm0.005\ $\tablefootmark{$\dagger$} \\[.6ex]
				$R_p/R_*$ & $0.0103\pm0.0005$ & $0.0505\substack{+0.0010\\-0.0009}$ \\[.6ex]
				$a/R_*$ & $54.4\substack{+1.2\\-1.4}\ $\tablefootmark{$\dagger$} & $29.0\substack{+0.5\\-0.7}\ $\tablefootmark{$\dagger$} \\[.6ex]
				Impact parameter & $0.00\pm0.15$ & $0.00\pm0.21$ \\[.6ex]
				Eccentricity & $0\ $\tablefootmark{*} & $0\ $\tablefootmark{*} \\[.6ex]
				Limb-darkening coefficients: & & \\
				\hspace{2em}$q_1=\left(u_1+u_2\right)^2$ & $0.08\substack{+0.15\\-0.06}$ & $0.65\substack{+0.22\\-0.23}$ \\[.3ex]
				\hspace{2em}$q_2=\frac{u_1}{2\left(u_1+u_2\right)}$ & $0.50\pm0.34$ & $0.43\substack{+0.22\\-0.16}$ \\[.6ex]
				\hline
			\end{tabular}
			\tablefoot{\tablefoottext{*}{Fixed parameter.}\tablefoottext{$\dagger$}{Unconstrained parameter distribution (posterior approximately equal to prior).}}
		\end{table*}
		
		We took advantage of this work to compare the performances of our tool and the version 10.1 of the data reduction pipeline of CHEOPS. The same analysis was performed with the light curves obtained with the DRP and we found residual noise levels of 9.1\,ppm for the Earth transit and 51.6\,ppm for the Neptune transit. We also compared the depths of the transits and found identical parameter values and uncertainties.

\section{Discussion and conclusion} \label{sec:conc}
	
	In this work, we first reported on CHEOPS performances measured during the payload calibration campaign. We focused on key-features and correction of instrumental effects to achieve ultra-high precision photometry. For instance, we showed that the characterisation of the photo-response non-uniformity of the detector allows us to perform flat-field correction for any given stellar spectrum with a precision of about 0.07\%. Then, we used the calibration products to carry out an end-to-end simulation. We analysed the simulated data and assessed mission performances to obtain precise stellar radius.
	
	The photometric precision of CHEOPS was measured by using long-term series of uniformly illuminated images acquired during the calibration campaign. After correction of instrumental effects, we extracted the average photometric signal of each exposure and corrected from calibration bench variations. Our analysis demonstrated that CHEOPS photometric performances when operated in a nominal condition are within requirements (20\,ppm over 6~hours).
	
	The data produced by \texttt{CHEOPSim} were analysed using an aperture photometry package developed specifically for this work. The results have been compared with the mission requirements. In the case of an Earth-size planet transiting a Sun-like star, we achieved a residual noise level of 10.2\,ppm over 6~hours, better than the results measured during the calibration campaign. The analysis of the transit fit suggests we can measure the planet-to-star radii ratio with a precision of 5\% , which corresponds to a SNR of 10 on the transit depth, compliant with the science objective of the mission. It is interesting to notice that, despite the interruption gaps, one reaches a very high SNR suggesting the detection may be possible with a single transit only. In the case of the Neptune-size planet in front of a K-dwarf star, we reached a residual noise of 51.7\,ppm over 3~hours, which is better than the photometric precision requirement of 85\,ppm. The precision on the planetary radius we obtained is 2\%, equivalent to a SNR of~25 on the transit depth. The corresponding SNR requirement is technically of 30 but does not account for interruptions in the light curves, which obviously reduce the number of data points and the overall precision. The correlation between the planetary radius and the impact parameter partially contribute to the increase of uncertainties.
	
	The results presented were based on measurements from the payload calibration and hence cover all the effects related to the instrument. This takes into account the chromatic transmission and the point spread function induced by the optical telescope, and the various effects linked to the detector and its read-out electronic chain (bias offset, dark current, gain and overall stability in time). Other aspects present once in space, such as the PSF shape in absence of gravity, the spatial distribution of the stray light or the variations of the telescope temperature along the orbit of CHEOPS, are expected to have a marginal impact on the global photometric error budget. In overall, all elements and measurements we collected during the pre-launch calibration activities suggest that CHEOPS is meeting all its requirements and demonstrates outstanding photometric performances of the system.

\begin{acknowledgements}\\
	This work has been carried out within the framework of the National Centre for Competence in Research PlanetS supported by the Swiss National Science Foundation. The authors acknowledge the financial support of the SNSF.\\
	The CHEOPS instrument and science simulator, \texttt{CHEOPSim}, is developed under the responsibility of the CHEOPS Mission Consortium. \texttt{CHEOPSim} is implemented by D.~Futyan as part of the Science Operation Centre located at the University of Geneva.\\
	We would also like to thank the referee for its contribution to this work.
\end{acknowledgements}

\bibliographystyle{aa}
\bibliography{references}

\end{document}